\documentclass[10pt]{article}
\usepackage{amsfonts}
\usepackage{mathrsfs}
\usepackage{amsmath}	
\usepackage{cite}
\usepackage{graphicx}
\usepackage{rotating}
\usepackage{hyperref}
\usepackage{verbatim}
\usepackage{slashed}
\usepackage[all]{xy}
\usepackage{xcolor}

\csname @addtoreset\endcsname{equation}{section}
\DeclareMathAlphabet\mathbfcal{OMS}{cmsy}{b}{n}

\newcommand{\beq}{\begin{equation}}
\newcommand{\eeq}{\end{equation}}
\newcommand{\bea}{\begin{eqnarray}}
\newcommand{\eea}{\end{eqnarray}}
\newcommand{\ba}{\begin{array}}
\newcommand{\ea}{\end{array}}
\newcommand{\bit}{\begin{itemize}}
\newcommand{\eit}{\end{itemize}}
\newcommand{\nn}{\nonumber}

\newcommand{\mezzo}{\frac{1}{2}}
\newcommand{\complesso}{{\ \hbox{{\rm I}\kern-.6em\hbox{\bf C}}}}
\newcommand{\reale}{{\hbox{{\rm I}\kern-.2em\hbox{\rm R}}}}
\newcommand{\uno}{ \,  \raisebox{+0.14em}{{\hbox{{\rm \scriptsize ]}} \raisebox{-0.2em}{\kern-.8em\hbox{1}}}} \, }  %  operatore identit\`a

\newcommand{\p}{\partial}
\renewcommand{\a}{\alpha}
\renewcommand{\b}{\beta}
\newcommand{\g}{\gamma}

\newcommand{\G}{\Gamma}
\renewcommand{\d}{\delta}
\newcommand{\D}{\Delta}

\newcommand{\Er}{{\mathbfcal{E}}}

\renewcommand{\k}{\kappa}
\renewcommand{\l}{\lambda}

\newcommand{\m}{\mu}

\newcommand{\n}{\nu}
\renewcommand{\r}{\rho}
\newcommand{\s}{\sigma}

\renewcommand{\t}{\theta}

\newcommand{\z}{\zeta}

\newcommand{\om}{\omega}

\begin{document}

%\begin{comment}

\begin{titlepage}
\begin{flushright}
%hep-th/??????\\
%UAI-PHY-19/01
IFUM-1080-FT
\end{flushright}
\vspace{1.7cm}
\begin{center}
\renewcommand{\thefootnote}{\fnsymbol{footnote}}
{\huge \bf Enhanced Ehlers Transformation and}
\vskip 6mm
{\huge \bf the Majumdar-Papapetrou-NUT Spacetime}
\vskip 31mm
{\large {Marco Astorino\footnote{Marco.Astorino@Gmail.com}}}\\
\renewcommand{\thefootnote}{\arabic{footnote}}
\setcounter{footnote}{0}
\vskip 10mm
{\small \textit{Istituto Nazionale di Fisica Nucleare (INFN), \\
 Sezione di Milano, Via Celoria 16, I-20133 Milano, Italy}
}
\end{center}
\vspace{4.5 cm}
\begin{center}
{\bf Abstract}
\end{center}
{The transformation which adds (or removes) NUT charge when it is applied to electrovacuum, axisymmetric and stationary space-times is studied. After analysing the Ehlers and the Reina-Treves transformations we propose a new one, more precise in the presence of the Maxwell electromagnetic field. The enhanced Ehlers transformation proposed turns out to act as a gravitomagnetic duality, analogously to the electromagnetic duality, but for gravity: it rotates the mass charge into the gravomagnetic (or NUT) charge.  \\
As an example the Kerr-Newman-NUT black hole is obtained with the help of this enhanced transformation. \\
Moreover a new analytical exact solution is built adding the NUT charge to a double charged black hole, at equilibrium. It describes the non-extremal generalisation of the Majumdar-Papapetrou-NUT solution. From the near-horizon analysis, its microscopic entropy, according to the Kerr/CFT correspondence, is found and the second law of black hole thermodynamics is discussed.}

\end{titlepage}

%\end{comment}

%\newpage

                        %%%%%%%%%%%%%%%%%%%%%%%%%%%%%%%%%%%%%%%%%%%%%%%%%%%%%%%%%%%%%%%%%%%%%%%
                        %                                                                     %
                        %                             START                                   %
                        %                                                                     %
                        %%%%%%%%%%%%%%%%%%%%%%%%%%%%%%%%%%%%%%%%%%%%%%%%%%%%%%%%%%%%%%%%%%%%%%%

\tableofcontents

\vspace{0.5 cm}

\section{Introduction}
\label{intro}
  
Recent gravitational waves detection confirms the existence of binary black holes systems. Actually because of the enormous emission of gravitational radiation provided by the interaction of black hole pairs, this will be the most abundant and relevant source of gravitational waves data in the near future.\\
Numerical relativity is developing some powerful tools to study black hole coalescence and merging. However analytical and exact solutions to qualitatively study phenomena involving an ensemble of interacting black holes are scarce. The few known belong to the class of static solutions, most of all supported by extra matter, such as cosmic strings, to sustain the gravitational attraction between the two sources and avoid the gravitational collapse, see for example \cite{israel-khan} or \cite{bonnor}. Actually, in the context of  static metrics, it have been shown that, under certain reasonable regularity assumptions, the only multi black hole configuration is an extremal ensemble of charged sources \cite{chrusciel}, \cite{hartle-hawking}. \\
On the other hand some hopes about a regular solutions may stem from the spin-spin repulsion effect of rotating bodies (eventually coupled with the repulsion electomagnetic effect). But because of the increasing technical complexity, the available stationary and rotating solutions involving more than one black hole, as the double Kerr-(Newman) solutions of \cite{neu} or \cite{dietz}, just to cite some famous cases, are even more rare than the static case. But generally these solutions are neither everywhere regular\footnote{For a very recent review of the problem see \cite{hennig}.} nor the two sources are causally connected, as in the case of accelerating and rotating metrics belonging to the Plebanski-Demianki family \cite{griffiths}. Between these kind of metrics the more physical are the ones where the conical singularities lie only between the two sources, avoiding to have conical defects extending to spacial infinity, which remains a globally well defined asymptotic region \cite{Cabrera-Munguia:2018omi} - \cite{manko-double-kerr-new}. \\
The purpose of this article is to enrich this scenario furnishing an analytical method to add rotation\footnote{The rotation we intend here does not refer to the angular momentum, such as the one possessed by the Kerr Black Hole. But we mean the rotation related to others orders in the angular multipole moments expansion, such the ones appearing in NUT solutions. Further details can be found in the section \ref{alekseev-beli-nut} and in the appendix \ref{app1}} to static solutions, including multi black hole ones, in the standard theory of general relativity, without the cosmological constant, coupled to Maxwell electromagnetism. \\
At this scope, in the context of axisymmetric and stationary space-times, some transformations known to add the Newman-Unti-Tamburino (NUT) parameter to a chosen seed solution, such as the Ehlers or the Reina-Treves transformation, are studied, in particular in section \ref{Ehlers-section}. In section (\ref{Kerr-Newman-NUT-sec}) we will see, throughout examples, how the known transformations of these kind, in the realm of pure general relativity, present some issues in the presence of the (standard) electromagnetic field.  \\
In section \ref{Ehlers-section} a new transformation able to add NUT charge to axisymmetric and stationary spacetimes, but without the criticalities of the known transformations, is proposed. In section \ref{sec-majumbdar} this enanched transformation is exploited to generate a stationary generalisation of the solution found by Alekseev and Belinski \cite{alekseev-belinski-4}, which is an analytical exact solution describing a couple of charged black hole at equilibrium. The extremal specialization of the Alekseev-Belinski-NUT spacetime reduce to the Majumdar-Papapetrou \cite{majumdar}-\cite{papapetrou} metric endowed with an extra NUT parameter. \\
The easy near horizon geometry of the latter solution allows one to follow the procedure provided by the Kerr/CFT correspondence \cite{strom08}, \cite{stro-duals}, \cite{strom-review}, \cite{compere-review} to address the issue of the microscopic entropy of the Majumdar-Papapetrou black holes.\\
Since these metrics can describe an ensemble of multiple gravitational sources endowed with  electromagnetic monopoles or, thanks to a specific choice of its parameters, a single charged source; it is possible to discuss the second law of black hole thermodynamics: which configuration is favoured from a thermodynamics point of view, specifically which of the two qualitative configurations is more likely to occur as a final state of a gravitational interaction. This is addressed in section \ref{sec-majumbdar}. \\

\section{Review of the Ernst generating technique}
\label{ernst-review-sec}

In this article we will focus on the standard theory of General Relativity coupled with Maxwell electromagnetism, governed by the following action
\beq  \label{action}
                       I[g_{\m\n}, A_\m] :=  \frac{1}{16 \pi G}  \int d^4x  \sqrt{-g} \left[ \textrm{R} - \frac{G}{\m_0} \ F_{\m\n} F^{\m\n} \right]  \ \ \ .
\eeq  
From (\ref{action}) it is possible to derive the field equations for the metric $g_{\mu \nu}$ and electromagnetic vector potential $A_\mu$
\bea  \label{field-eq-g}
                        &&   \textrm{R}_{\m\n} -   \frac{\textrm{R}}{2}  g_{\m\n} = \frac{2G}{\m_0} \left( F_{\m\r}F_\n^{\ \r} - \frac{1}{4} g_{\m\n} F_{\r\s} F^{\r\s} \right)  \quad ,   \\
       \label{field-eq-A}                  &&   \partial_\m ( \sqrt{-g} F^{\m\n}) = 0  \ \quad . 
                        \eea
The electromagnetic Faraday tensor $F_{\mu\nu}$ is defined, as usual, from the $U(1)$ gauge four-potential $F_{\mu\nu}:=\p_\mu A_\nu - \p_\nu A_\mu$. 
The most generic axisymmetric and stationary spacetime, containing  two commuting Killing vectors $\p_t$ and $\p_\varphi$, can be written, for this theory, in the Lewis-Weyl-Papapetrou (LWP) form as
\beq \label{lwp-metric}
                         ds^2 = - f \left( dt + \om d\varphi \right)^2 + f^{-1} \left[ \rho^2 d\varphi^2 + e^{2\gamma}  \left( d \rho^2 + d z^2 \right) \right] \ .
\eeq
All the three structure functions appearing in the metric $f,\omega$ and $\gamma$ depends only on the non-Killing coordinates $(\rho, z)$. We will consider a generic electromagnetic potential compatible with the spacetime symmetries, and the circularity of the LWP metric, given by  $A=A_t(\rho, z) dt + A_\varphi(\rho,z) d\varphi$. For a discussion about the generality of the electromagnetic ansatz see \cite{carter-meridional}. \\
Ernst in \cite{ernst2} discovered that, when the equations of motion (\ref{field-eq-g})-(\ref{field-eq-A}) are restricted to the above axisymmetric and stationary ansatz, they reduce to a couple of complex vectorial differential equations, as follow\footnote{Henceforward in the paper the ratio between the Newton constant and the vacuum permeability is fixed $G/\m_0=1$ without loosing generality.}  
\bea 
     \label{ee-ernst}  \left( \textsf{Re} \ \Er + | \mathbf{\Phi} |^2 \right) \nabla^2 \Er   &=&   \left( \overrightarrow{\nabla} \Er + 2 \ \mathbf{\Phi^*} \overrightarrow{\nabla} \mathbf{\Phi} \right) \cdot \overrightarrow{\nabla} \Er   \quad ,       \\
     \label{em-ernst}   \left( \textsf{Re} \ \Er + | \mathbf{\Phi} |^2 \right) \nabla^2 \mathbf{\Phi}  &=& \left( \overrightarrow{\nabla} \Er + 2 \ \mathbf{\Phi^*} \overrightarrow{\nabla} \mathbf{\Phi} \right) \cdot \overrightarrow{\nabla} \mathbf{\Phi} \quad ,
\eea
and two other first order partial differential equations for $\gamma(\r,z)$,  decoupled from the previous ones, (\ref{ee-ernst}) and (\ref{em-ernst})
\bea
   \label{gamma-rho}       \p_\r \gamma(\r,z)  &=&  \frac{\r}{4 [Re(\Er) +\mathbf{\Phi}\mathbf{\Phi}^*]^2} \bigg[\Big( \p_\r \Er + 2 \mathbf{\Phi}^*\p_\r \mathbf{\Phi}\Big) \Big( \p_\r \Er^* + 2 \mathbf{\Phi}\p_\r \mathbf{\Phi}^* \Big) - \Big( \p_z \Er + 2 \mathbf{\Phi}^*\p_z \mathbf{\Phi}\Big) \Big( \p_z \Er^* + 2 \mathbf{\Phi}\p_z \mathbf{\Phi}^* \Big)  \bigg] \nn \\
                                                &-& \ \frac{\r}{Re(\Er) +\Phi\mathbf{\Phi}^*} \quad \ \Big( \p_\r \mathbf{\Phi}\p_\r\mathbf{\Phi}^* -\p_z \mathbf{\Phi}\p_z\mathbf{\Phi}^* \Big)    \qquad  ,\\
    \label{gamma-z}         \p_z \gamma(\r,z)  &=&   \frac{\r}{4 [Re(\Er) +\mathbf{\Phi}\mathbf{\Phi}^*]^2} \bigg[\Big( \p_\r \Er + 2 \mathbf{\Phi}^*\p_\r \mathbf{\Phi}\Big) \Big( \p_z \Er^* + 2 \mathbf{\Phi}\p_z \mathbf{\Phi}^* \Big) + \Big( \p_z \Er + 2 \mathbf{\Phi}^*\p_z \mathbf{\Phi}\Big) \Big( \p_\r \Er^* + 2 \mathbf{\Phi}\p_\r \mathbf{\Phi}^* \Big)  \bigg] \nn  \\
                                                &-& \ \frac{\r}{Re(\Er) +\mathbf{\Phi}\mathbf{\Phi}^*} \quad  \ \Big( \p_\r \mathbf{\Phi}\p_z\mathbf{\Phi}^* + \p_z \mathbf{\Phi}\p_\r\mathbf{\Phi}^* \Big)  \qquad .
\eea
The complex Ernst potential are defined as 
\beq \label{def-Phi-Er} 
       \mathbf{\Phi} := A_t + i \tilde{A}_\varphi  \qquad , \qquad \qquad     \Er := f - \mathbf{\Phi} \mathbf{\Phi}^* + i h  \quad ,
\eeq
where $\tilde{A}_\varphi$ and $h$ can be obtained from
\bea
    \label{A-tilde} \overrightarrow{\nabla} \tilde{A}_\varphi &:=& - f \r^{-1} \overrightarrow{e}_\varphi \times (\overrightarrow{\nabla} A_\varphi - \omega  \overrightarrow{\nabla} A_t ) \\
    \label{h}    \overrightarrow{\nabla} h &:=& - f^2 \r^{-1} \overrightarrow{e}_\varphi \times \overrightarrow{\nabla} \omega - 2 \ \textsf{Im} (\mathbf{\Phi}^*\overrightarrow{\nabla} \mathbf{\Phi} )  
\eea
 
So the Ernst equations constitute the main equations for the physical system, because once the Ernst potential, satisfying   (\ref{ee-ernst}) and (\ref{em-ernst}), are known, $\gamma(\r,z)$ is obtained by quadratures from (\ref{gamma-rho}) - (\ref{gamma-z}).\\
The differential operators ($\overrightarrow{\nabla}$ and $\nabla^2$) appearing in (\ref{ee-ernst})-(\ref{em-ernst}) are just the flat gradient and Laplacian in cylindrical Weyl	 coordinates ($\r,z,\varphi$)\footnote{In appendix \ref{app2} some notation on differential operator in various coordinates can be found.}.
 
The equations of motion for the stationary and axisymmetric complex Ernst potentials (\ref{ee-ernst}) and (\ref{em-ernst}) can be deduced from the following effective action for the complex fields couple ($\Er, \mathbf{\Phi}$) 

\beq \label{ernst-action}
        I(\Er,\mathbf{\Phi})= \int dz \int d\r \left[\frac{ \big( \overrightarrow{\nabla} \Er +2\mathbf{\Phi}^*  \overrightarrow{\nabla} \mathbf{\Phi}\big) \big( \overrightarrow{\nabla} \Er^* +2\mathbf{\Phi} \overrightarrow{\nabla} \mathbf{\Phi}^* \big)}{\big(\Er+\Er^*+2\mathbf{\Phi}\mathbf{\Phi}^*\big)^2} - \frac{  \overrightarrow{\nabla} \mathbf{\Phi} \overrightarrow{\nabla} \mathbf{\Phi}^*}{\Er+\Er^*+2\mathbf{\Phi}\mathbf{\Phi}^*} \right]
\eeq

From the above Lagrangian density in the square brackets, it's possible to derive that the Ernst equations for the complex fields ($\Er, \mathbf{\Phi}$) have some remarkable non-trivial Lie point symmetries properties \cite{stephani}, \cite{marcoa-lambda} which form the $SU(2,1)$ group. These symmetries can be written as a set of five independent transformation which leave invariant the action (\ref{ernst-action}) and its equation of motion (\ref{ee-ernst})-(\ref{em-ernst}):
\bea \label{su21-transf}
      (I)    && \Er \longrightarrow \Er' = \l \l^* \Er  \qquad \ \quad \qquad \ ,  \qquad \mathbf{\Phi} \longrightarrow  \mathbf{\Phi}' = \l \mathbf{\Phi} \quad , \nn \\
      (II)   && \Er \longrightarrow \Er' = \Er + i \ b \qquad  \ \ \ \qquad, \qquad \mathbf{\Phi} \longrightarrow  \mathbf{\Phi}' = \mathbf{\Phi} \quad ,  \nn \\
      (III)  && \Er \longrightarrow \Er' = \frac{\Er}{1+ic\Er} \qquad \ \ \qquad , \qquad  \mathbf{\Phi} \longrightarrow  \mathbf{\Phi}' = \frac{\mathbf{\Phi}}{1+ic\Er} \quad ,  \\
      (IV)   && \Er \longrightarrow \Er' = \Er - 2\b^*\mathbf{\Phi} - \b\b^* \ \ \ , \qquad \mathbf{\Phi} \longrightarrow  \mathbf{\Phi}' = \mathbf{\Phi} + \b  \quad ,  \nn \\
      (V)    && \Er \longrightarrow \Er' = \frac{\Er}{1-2\a^*\mathbf{\Phi}-\a\a^*\Er} \  , \quad  \ \ \mathbf{\Phi} \longrightarrow  \mathbf{\Phi}' = \frac{\mathbf{\Phi}+\a\Er}{1-2\a^*\mathbf{\Phi}-\a\a^*\Er}\ \quad  \nn
\eea
where $b, c \in \mathbb{R}$ and $\a, \l,\b \in \mathbb{C}$. Some of these transformation are just gauge symmetries and can be reabsorbed by a coordinate transformation, while others actually have non-trivial physical effects. The combination of (I)-(V) generate other transformation, for example applying (I)-(III) in a certain limit of the parameters\footnote{ \label{foot-inv} It is easy to verify that the limit for $b\rightarrow\infty$ of the product of the transformations $(I)\circ(III)\circ(II)$ for $c=b^{-1}$ and $\l=ib^{-1}$ leads to (\ref{inv}) \cite{stephani}.} gives the {\it inversion} transformation 
\beq  \label{inv}
      (inv) \qquad \quad   \Er  \longrightarrow \Er' = \frac{1}{\Er}   \quad , \qquad  \mathbf{\Phi} \longrightarrow  \mathbf{\Phi}' = \frac{\mathbf{\Phi}}{\Er}  \quad ;
\eeq
which will be useful in the next sections.  A particular specialization, for null electromagnetic field, of this inversion transformation is known as the Buchdahl transformation.  \\
In this article we will mainly focus on the transformation of type $(III)$ called the Ehlers transformation. It maps a given solution of the axisymmetric and stationary Einstein-Maxwell equations, identified by the Ernst potentials $(\Er,\mathbf{\Phi})$, in another non-equivalent one $(\Er',\mathbf{\Phi}')$. The Ehlers transformation is parametrised by a real number $c$ which introduce an extra parameter to the seed solution usually interpreted as the NUT (Newman-Unti-Tamburino) charge\footnote{The term charge is possibly abused in this context because the NUT parameter is not actually related to a conserved quantity associated to a physical symmetry, as it occurs for the mass or the angular momentum. But still this terminology is often used in the literature, usually referring to a topological invariant quantity that can be associated to the NUT parameter, similarly to the magnetic monopole charge \cite{nicolai-stelle}. In fact, from the analogy with the duality between the electric and magnetic monopoles charges, the NUT parameter is usually denoted as the gravitomagnetic monopole mass, as the dual to the standard gravitational monopole mass. This because the mass parameter mainly contributes to the electric part of the Weyl tensor decomposition, while the NUT parameter to the magnetic sector. Physically this property is related to the fact that the NUT parameter introduces an interaction on trajectories of massive test particles similar to the magnetic force on a charged particle, even if it is a purely gravitational solution. For this reason sometimes in the literature the NUT parameter is associated to {\it radial} angular momentum to distinguish it from standard {\it axial} angular momentum of the Kerr black hole.}.\\
Recently the physical significance of the NUT generalisation of lorenzian black holes solutions in general relativity has been partially rehabilitated, under certain assumptions \cite{clement-rehab}. Nevertheless there is a open discussion about the physical interpretation of the singularities, that may take place in the presence of the NUT parameter. In fact the nodal singularities typically appearing on the axis of symmetry, usually called Misner strings, can be removed by a periodic time identification, which naturally generates closed timelike curves. Therefore some people prefers to keep the string and interpret it as a singular material source of angular momentum.  \\ 
In \cite{reina-treves} the Ehlers transformation, although in another form, was applied to the Kerr black hole seed to obtain the Kerr-NUT spacetime. However we will make use of the Ehlers transformation in the form of (\ref{su21-transf}-III) because it is easier to apply it to generic Ernst seed potentials.

\section{Example: adding NUT to the Kerr-Newman black hole}
\label{Kerr-Newman-NUT-sec}

In this section, as an example to show how the solution generating tecnhique works, we will generalise the work of  \cite{reina-treves} to show how to obtain the Kerr-Newman-NUT solution from the Kerr Newman black hole.\\
The Kerr-Newman spacetime describes a rotating and charged asymptotically flat black hole in the theory of general relativity. In the presence of both electric and magnetic monopole charge, respectively labelled $q$ and $p$, the solution can be written in Boyer-Lindquist coordinates as follows

\bea \label{kerr-new-metric}
ds^2 &=& - \frac{\Delta(r)}{R(r,\theta)^2} \left( dt - a \sin^2 \theta \ d\varphi \right)^2  + R(r,\theta)^2 \left( \frac{dr^2}{\Delta(r)} + d\theta \right) + \frac{\sin^2 \theta}{R(r,\theta)^2} \left[ a dt -(r^2+a^2) d\varphi \right]^2 \quad , \qquad   \\
A_\mu &=&  \left[ \frac{qr+pa\cos \theta}{R(r,\theta)^2}  \ , 0 \ ,  0 \ ,  \frac{p \cos \theta (r^2+a^2) - aqr \sin^2 \theta %+ (qr+pa\cos \theta) \omega_0 
}{R(r,\theta)^2} \right] \quad , \qquad\label{kerr-new-potential}
\eea
where
\beq
            R(r,\theta) := r^2 + a^2 \cos^2 \theta   \qquad , \qquad \qquad  \Delta(r):= r^2 - 2 m r +a^2+q^2+p^2  \quad . 
\eeq

The first step of the Ernst procedure consists in identifying the $f,\omega,\gamma$ functions appearing in the Lewis-Weyl-Papapetrou ansatz (\ref{lwp-metric}) for the seed metric, which in this case is the Kerr-Newman one (\ref{kerr-new-metric}). % in spherical coordinates (r,x). Then 
At this purpose the coordinates transformation 

\beq
               \rho(r,\theta):= \sin \theta \sqrt{\D(r)} \qquad , \qquad \qquad   z(r,\theta):=\cos \theta (r-m)
\eeq

is applied to (\ref{lwp-metric}) to get a better suited LWP line element for the seed coordinates

\beq \label{lwp-r-theta}
               ds^2 = - f(r,\theta) \left[ dt + \om(r,\theta) d\varphi \right]^2 + \frac{1}{f(r,\theta)} \left\{ e^{2\gamma(r,\theta)} \left[ (r-m)^2-\k^2 \cos^2\theta \right] \left[ \frac{d r^2}{\Delta(r)} + d \theta^2 \right] + \sin^2 \theta \ \Delta(r) d\varphi^2  \right\}   .
\eeq
The constant $\k$, specifically for the Kerr-Newman spacetime, takes the value $\k=\sqrt{m^2-a^2-q^2-p^2}$.\\ 
By comparing the metrics (\ref{kerr-new-metric}) and (\ref{lwp-r-theta}) it is possible to determinate the structure functions of the Kerr-Newman metric. We will present them in more ergonomic coordinates ($r,x:=\cos \theta$):
\bea
     \label{fkn}    f_0(r,x) &=& 1 + \frac{q^2+p^2-2mr}{r^2+a^2x^2} \\
      \label{wkn}   \omega_0(r,x) &=& \frac{a(1-x^2)(2mr-q^2-p^2)}{2(r^2-2mr+q^2+p^2+a^2x^2)}    \\
       \label{Gkn}    e^{2\gamma_0}(r,x)  &=&   \frac{r^2-2mr+q^2+p^2+a^2x^2}{(r-m)^2-\k^2 x^2}
\eea
The differential operators $\overrightarrow{\nabla}$ and $\nabla^2$ in terms of the coordinates ($r,x$) becomes
\bea
       \label{gradient}    \overrightarrow{\nabla} \phi(r,x) = \frac{1}{\sqrt{(r-m)^2-\k^2x^2}} \left[ \overrightarrow{e}_r \sqrt{\D(r)} \ \frac{\p \phi(r,x)}{\p r}  + \overrightarrow{e}_x \sqrt{1-x^2} \ \frac{\p \phi(r,x)}{\p x} \right]  \\ 
  \label{laplacian}         \nabla^2 \phi(r,x) =\frac{1}{(r-m)^2-\k^2x^2}\left\{ \frac{\p}{\p r} \left[ \D(r) \frac{\p \phi(r,x) }{\p r} \right] + \frac{\p}{\p x} \left[ (1-x^2) \frac{\p \phi(r,x) }{\p x} \right] \right\}
\eea
The ($r,x$) coordinates are closely related with the prolate spherical ones ($y,x$). To obtain these latter is sufficient to define $y:=(r-m)/\k$. \\
Then in order to identify the electromagnetic seed Ernst potential $\mathbf{\Phi}_0$\footnote{The zero subscript in $\mathbf{\Phi}_0$ point out that we refer to the seed fields.}, as defined in (\ref{def-Phi-Er}), for the Kerr-Newman gauge field we need to derive $\tilde{A}_\varphi$ from eq (\ref{A-tilde}) and taking into account eqs. (\ref{kerr-new-potential}), (\ref{fkn})-(\ref{gradient}). For the seed under consideration we have
\beq \label{KN-seed-Phi0}
     \tilde{A}_{\varphi 0} (r,x) =   \frac{a q x - p r}{r^2 + a^2 x^2}  \qquad      \Longrightarrow  \qquad  \mathbf{\Phi}_0 (r,x) = - \frac{q+ ip}{r+ i a x} \quad .
\eeq 
While to obtain the gravitational Ernst potential for the seed Kerr-Newman metric $\Er_0$ we have first to integrate (\ref{h}) to get
\beq \label{KN-seed-Er0}
       h_0(r,x) =  \frac{2 m x}{r^2+a^2 x^2}  \qquad      \Longrightarrow  \qquad  \Er_0 (r,x) =  1 - \frac{2 m}{r + i a x} \quad .
\eeq   
Now we can apply the Ehlers transformation  (\ref{su21-transf}-III) to the seed Ernst potentials ($\Er_0,\mathbf{\Phi}_0$) to generate a new axisymmetric and stationary solution of the Einstein-Maxwell theory in terms of the Ernst potentials
\bea \label{ehlers_Er1}
        \Er(r,x) &=&  \frac{\Er_0 (r,x)}{1+ic\Er_0 (r,x)}  \ = \  \frac{ax +2 i m - ir}{ax-ir +c(r-2m+iax)}  \quad , \\
         \mathbf{\Phi}(r,x) &=&  \frac{\mathbf{\Phi}_0 (r,x)}{1+ic\Er_0 (r,x)} \ = \ \frac{-p+iq}{ax-ir +c(r-2m+iax)}            \quad . \label{ehlers_Phi1}
\eea
To come back to the metric and vector potential representation it is sufficient to use the definitions (\ref{def-Phi-Er}) - (\ref{h}). In particular $f, \om, A_t$ can be read directly from (\ref{def-Phi-Er})
\bea
       h (r,x) &=&  \textsf{Im}(\Er) \ = \ \ \frac{2amx -c[(r^2-2m)^2+a^2x^2]}{r^2-4acmx+a^2x^2+c^2[(r^2-2m)^2+a^2x^2]} \quad ,\quad  \\
       f (r,x) &=& \textsf{Re}(\Er) + \mathbf{\Phi}\mathbf{\Phi}^*   \ = \ \frac{p^2+q^2-2mr+r^2+a^2x^2}{r^2-4acmx+a^2x^2+c^2[(r^2-2m)^2+a^2x^2]} \quad , \quad \label{f2} \\
       A_t (r,x) &=& \textsf{Re}(\mathbf{\Phi}) \ = \ \frac{-qr-apx+c(2mp-pr+aqx)}{r^2-4acmx+a^2 x^2+c^2[(r^2-2m)^2+a^2x^2]} \quad , \quad \label{Atpredual} \\
       \tilde{A}_\varphi (r,x) &=& \textsf{Im}(\mathbf{\Phi}) \ = \ \frac{-2cmq-pr+cqr+a(cp+q)}{r^2-4acmx+a^2 x^2+c^2[(r^2-2m)^2+a^2x^2]} \quad . \quad 
\eea
The above quantities have to be inserted in the equation (\ref{h}) to obtain 
\beq \label{om2}
\om(r,x) =  -4cmx + \frac{a(x^2-1)\left\{p^2+q^2-2mr+c^2\left[p^2+q^2+2m(r-2m) +4acmx\right]\right\}}{r^2-2mr+q^2+p^2+a^2x^2}  + \om_0 \quad .\ \
\eeq
Then, finally, also (\ref{A-tilde}) can be solved to get 
\beq \label{Af2}
A_\varphi(r,x) =  px-cqx - \frac{[-2cmp+cpr+qr+a(p-cq)x][-4cmx+a(1+c^2)(x^2-1)+\om_0]}{r^2-4acmx+a^2x^2+c^2[(r-2m)^2+a^2x^2]} + A_{\varphi0} \quad .
\eeq
The arbitrary constants $\om_0$ and $A_{\varphi0}$ usually are constrained by asking regularity of the metric and the magnetic field on the symmetry axis $\rho=0$. In particular the magnetic field, to be globally well behaved, should fulfil the physical requirement assuring that the magnetic monopole moments must be null on the axis of symmetry \cite{carter}, therefore
\beq
           \lim_{\rho \rightarrow 0} A_{\varphi}(\rho,z) = 0      \quad  . 
\eeq
The Ehlers transformation is not affecting the $\g$ function, which remains the same as the seed $\g_0$. This can  be directly verified, substituting (\ref{ehlers_Er1})-(\ref{ehlers_Phi1}) into eqs. (\ref{gamma-rho}) - (\ref{gamma-z}).  \\
The solution generated by the Ehlers transformation represents a Kerr-Newman black hole in a Taub-NUT background, whose electromagnetic vector potential is generally written in this form
\beq  \label{hatA}
      \hat{A}_\mu =   \left\{ - \frac{q r +p (ax+\ell)}{r^2+(\ell+ax)^2} ,\ 0 , \ 0 , \ -q \ \frac{r(x-1)(a+ax+2\ell)}{r^2+(\ell+ax)^2} +p \frac{(ax+\ell)[r^2+(a+\ell)^2]}{a [r^2+(\ell+ax)^2] } -\hat{\om}_0 \hat{A}_t(r,x) + \hat{A}_{\varphi0} \right\} \ .
\eeq

To have a well defined Maxwell potential, in the no rotation limit, the constant $\hat{A}_{\varphi0}$ have to be fiexd to $-\frac{p\ell}{a}$. \\
In the case one wants to verify the equivalence of the generated metric and the standard Kerr-Newman-NUT one, denoted as $d\hat{s}^2$, it is necessary a coordinate transformation and a rescaling some parameters.\\
Interesting enough the same procedure is not sufficient to get the electromagnetic potential $\hat{A}_\mu$ of eq. (\ref{hatA}), in fact neither the asymptotic behaviour of the RN-NUT electromagnetic field is retrieved. A further duality transformation on the electromagnetic field is required. Of course in four-dimensions this transformation is not affecting the metric. In terms of the Ernst potential the duality transformation can be written as
\beq \label{dual-trasf}
        \mathbf{\mathbf{\Phi}}\longrightarrow \bar{\mathbf{\Phi}}= \mathbf{\Phi}\exp(i \b) \quad .
\eeq 
Actually this is a special unitary sub-case, when $\l=\exp (i \b)$, of the more general symmetry (\ref{su21-transf})-$(I)$ of the Ernst electrovacuum equations (\ref{ee-ernst})-(\ref{em-ernst}).\\ 
%\beq
%       \Er \longrightarrow \bar{\Er} = \l \l^* \Er \quad , \qquad \    \mathbf{\Phi}\longrightarrow \bar{\mathbf{\Phi}}= \l \mathbf{\Phi} \quad , \qquad \l \in \mathbb{C}             \quad .
%\eeq
Thus the non-null components of the rotated electromagnetic vector potential (\ref{Atpredual}) and (\ref{Af2}) can be obtained by (\ref{def-Phi-Er}) and (\ref{A-tilde}). They become respectively
\bea \label{barAt}
         \bar{A}_t (r,x) &=& \frac{-\left[ -2 c m p + c p r +q r + a x (p-c q)  \right] \cos \b + \left[2 c m q +pr -c q r -ax (cp+p) \right] \sin \b }{r^2-4 a c m x + a^2 x^2 + c^2 \left[ (r-2m)^2 +a^2 x^2 \right]}  \quad , \qquad \\
      \label{barAf}   \bar{A}_\varphi (r,x) &=&  (p-cq) x \cos \b + (c p + q) x \sin \b - [-4 cmx +a (1+c^2)(x^2-1) +\omega_{0} ] \bar{A}_t (r,x) \quad  .  \qquad        
\eea 
To check the equivalence of the electromagnetic potentials (\ref{barAt})-(\ref{barAf}) and (\ref{hatA}) the following coordinates transformation of the temporal and radial coordinate is needed\footnote{$\hat{\om}_0$ and$\hat{A}_{\varphi0}$ are usually chosen to be null, while $\om_0=-2\ell$.}
\beq
       r \longrightarrow \bar{r} = r\sqrt{1+c^2}  - \frac{2 \ c^2 m}{\sqrt{1+c^2}}  \qquad , \qquad \qquad t \longrightarrow \bar{t}= \frac{t}{\sqrt{1+c^2}} \quad ,
\eeq
and a rescaling of the physical parameters\footnote{This parametrization is consistent for the negative branch of $\ell$, when $\ell>0$ some signs change.} 
\beq
        q  \longrightarrow \bar{q} = q\sqrt{1+c^2} \quad , \qquad   p \longrightarrow \bar{p} =  p \sqrt{1+c^2} \quad ,
\eeq  
\beq
\ \ \ \        a \longrightarrow \bar{a} = a  \sqrt{1+c^2} \quad , \qquad  m \longrightarrow \bar{m} = -\frac{\ell \sqrt{1+c^2}}{2c} \quad .
        %= \frac{1}{\sqrt{2}} \sqrt{\ell^2+m\left( m+\sqrt{m^2+\ell^2} \right)}  .
\eeq
Where the relation between the extra parameter introduced by the Ehlers transformation $c$  and the standard NUT one $\ell$ is 
\beq
         \quad , \qquad   c = \frac{m-\sqrt{m^2+\ell^2}}{\ell^2} \quad .
\eeq  
Finally the particular duality rotation (\ref{dual-trasf}), which completes the alignment of the electromagnetic vector potentials $\bar{A}_\m$ and $\hat{A}_\m$,  is given by
\beq
          \cos \b =  \frac{1}{\sqrt{1+c^2}} = \sqrt{\frac{1}{2}+\frac{m}{2\sqrt{m^2+\ell^2}}} \quad .
\eeq
Thanks to the above coordinates transformation also the Ehlers transformed Kerr-Newman metric, basically determined by eqs. (\ref{lwp-metric}), (\ref{Gkn}), (\ref{f2}) and (\ref{om2}), fits into the usual dyonic Kerr-Newmann-NUT form 
\beq \label{kn-nut-standard-inizio}
       \hat{ds}^2 = - \hat{f} \left( dt + \hat{\om} d\varphi \right)^2 + \hat{f}^{-1} \left[ \alpha^2 d\varphi^2 + e^{2\hat{\gamma}}  \left( \frac{d r^2}{Q(r)} + \frac{d x^2}{P(x)} \right) \right] \ .
\eeq
where
\bea
       \hat{f}(r,x)   &=&               \frac{Q(r)-a^2P(x)}{r^2+(\ell+ax)^2}  \quad ,  \\
       \hat{\om}(r,x) &=&  -\frac{a[r^2+(a+l)^2P(x)]+(x-1)(a+ax+2\ell)Q(r)}{a^2P(x)-Q(r)} - \hat{\om}_0         \quad ,        \\
       \a(r,x)  &=&  \sqrt{Q(r)P(x)}     \quad ,              \\ 
       \hat{\g}(r,x)  &=&  Q(r)-a^2P(x)     \quad ,               \\
       Q(r)     &=&  r^2 - 2mr +a^2+q^2+p^2-\ell^2    \quad ,                 \\
       P(x)     &=&  1-x^2     \quad . \label{kn-nut-standard-fine}
\eea
This procedure is completely generic and does not apply only to the Kerr-Newman spacetime, but to all axisimmetric and stationary spacetimes of Einstein-Maxwell theory. \\
In the section \ref{sec-majumbdar} the same technique will be exploited to obtain a new solution describing a NUT generalisation of a couple of charged black holes, but through a slightly different Ehlers transformation, which will be described in the next section. \\

\section{Enhanced Ehlers transformations and gravitomagnetic duality}
\label{Ehlers-section}
 
% \begin{figure}[h!]
%  \centering
% \includegraphics[scale=0.7]{rprA.pdf} 
%  \caption{\small Positions of the Killing horizons for a fixed value of the mass parameter ($m=5$). The picture does not change qualitatively for different values of m or range of $E$ and $q$.}
%\label{raggi} 
% \end{figure} 

Originally Ehlers, in his 1959 thesis, discovered a symmetry of the Ernst field equations (\ref{ee-ernst})-(\ref{em-ernst}) that can be written as follows
\beq \label{original-ehlers} \left\{\begin{matrix}
 U + W  &\longrightarrow &  U + W \ + \  i \ c \ (U-W) \\
  V &\longrightarrow & V \\ 
U-W &\longrightarrow & U-V
\end{matrix}\right.
\eeq

for any complex Ernst potential of the form

\beq
            \Er = \frac{U-W}{U+W}      \quad , \qquad \mathbf{\Phi}= \frac{V}{U+W}  \quad .
\eeq

It easy to show that the (\ref{original-ehlers}) transformation is equivalent to the Ehlers transformation written in the form of eq (\ref{su21-transf}-III). Nevertheless in the previous section we have realised that, in general, the so called Ehlers transformation given in eqs (\ref{su21-transf}-III) is not sufficient to add the NUT charge to the electrovacuum axisymmetric spacetime seed. When the seed presents a non-null Maxwell electromagnetic field, the Ehlers transformation (\ref{su21-transf}-III) adds an additional (and undesired) rotation to the $U(1)$ electromagnetic vector potential and an extra duality transformation of the Ernst potential is needed. \\
On the other hand Reina and Treves in \cite{reina-treves} pointed out how to add NUT charge to spacetimes whose Ernst complex potentials can be written as

\beq \label{Er-Phi-asym-flat}
       \Er = \frac{\xi-1}{\xi+1}       \quad , \qquad \mathbf{\Phi}=\frac{q}{\xi+1}     \quad .
\eeq

As explained by Ernst in \cite{ernst2}, the above form for the complex potentials stems, from the original one, as defined in (\ref{def-Phi-Er}), requiring that $\Er$ is an analytic function of $\mathbf{\Phi}$ and using boundary conditions which mimic the asymptotic flatness: $\Er=1$ and $\mathbf{\Phi}=0$ at spatial infinity. In fact, as the Kerr-Newman solution fulfil these requirements , it can be cast into the form (\ref{Er-Phi-asym-flat}). Actually in \cite{reina-treves} only uncharged solutions were treated such as the Kerr black hole or the Tomimatsu-Sato metric. Anyway it can be shown that the NUT parameter can be added to the seed spacetime written as (\ref{Er-Phi-asym-flat}), just rotating the complex function $\xi$ by a constant phase in the complex plane
\beq    \label{reina-tr}  
                             \xi \longrightarrow \bar{\xi} = \xi e^{i\tau} \quad .
\eeq
However in presence of the electromagnetic potential, just as for the Ehlers transformation, the procedure is not so straight. \\
First of all we would like to clarify the relation between the Ehlers transformation and the Reina-Treves one (\ref{reina-tr}), apart the mentioned reduced domain of applicability of the latter. Applying the following transformations to the complex Ernst potentials (\ref{Er-Phi-asym-flat})
\beq \label{reina-comp}
       (inv) \circ (II) \circ  (I) \circ (inv) \circ (II)  \circ \ \left(\begin{matrix}
 \Er  \\
 \mathbf{\Phi} 
\end{matrix} \right) \  =  \   \left(\begin{matrix}
 \bar{\Er}  \\
 \bar{\mathbf{\Phi}}  
\end{matrix} \right) \quad ,
\eeq
and considering  $\l=1-ib$ we get the transformed complex potentials
\beq
 \bar{\Er} = \frac{\xi(1+ib)-(1-ib)}{\xi(1+ib)+(1-ib)}  \quad , \qquad  \bar{\Phi} = \frac{q(1-ib)}{\xi(1+ib)+(1-ib)} \quad .
\eeq
Finally defining $\tau := \arccos \left( \frac{1-b^2}{1+b^2} \right) $ the above Ernst potentials transform in 
\beq \label{reina-treves-trasf}
         \bar{\Er} = \frac{\xi e^{i\tau}-1}{\xi e^{i\tau}+1}  \quad , \qquad  \bar{\mathbf{\Phi}} = \frac{q}{\xi e^{i\tau}+1} \quad ,
\eeq
which exactly correspond to the transformation given in (\ref{reina-tr}), when applied to the potentials of the form (\ref{Er-Phi-asym-flat}). Therefore we have shown how the Reina-Treves transformation can be deduced from a proper composition of the basic invariance symmetries (\ref{su21-transf}) of the Ernst Fields  equations. Note that the relation between the Reina-Treves and Ehlers transformation relies on the fact that, the latter is a part of the inverse transformation we used in eq (\ref{reina-comp}), as explained in footnote \ref{foot-inv}.     \\ 
Now that we know how to obtain the Reina-Treves transformation from the basic $SU(2,1)$ symmetries transformations  (\ref{su21-transf}), we can extend the transformation (\ref{reina-tr}) for more general Ernst potentials with respect to the asymptotic flat subclass (\ref{Er-Phi-asym-flat}) of the Reina-Treves. Just applying the sequence of transformations (\ref{reina-comp}) to unconstrained Ernst potentials we get
\beq \label{gen-reina-treves}
          \Er \longrightarrow \bar{\Er} = \frac{\Er+ib}{1+ib\Er} \qquad \ \ , \ \ \qquad \mathbf{\Phi}\longrightarrow \bar{\mathbf{\Phi}} = \frac{\mathbf{\Phi}(1-ib)}{1+ib\Er} \quad .
\eeq 
Note that, when the Reina-Treves transformation is written in this generalised form, the similarities with the Ehlers transformation are quite apparent.\\ 
Nevertheless, repeating the procedure discussed in the previous section for the Ehlers transformation, but this time using the transformation proposed by Reina-Treves, we realise that also in this case an extra duality transformation is needed in the presence of the electromagnetic field (followed by a coordinates transformation\footnote{Note that coordinate transformations and the symmetries of the Ernst potentials (\ref{su21-transf}) do \underline{not} commute in general. Therefore, in order to have some control on the resulting spacetime, when several consecutive transformations are composed it is better to avoid coordinate transformations.} too).\\
Instead it would be optimal to know exactly which is the transformation able to \underline{only} add the NUT charge to a chosen seed spacetime. This point turns out to be particularly relevant, as in the next section case, where the result of the transformation is an unknown solution, thus in principle we ignore if  extra manipulations or transformations are needed to get only the NUT extension (and eventually which ones).\\
After having repeated the previous section algorithm to add the NUT charge to the Kerr-Newman black hole for the Reina-Treves transformation (\ref{reina-treves-trasf}) or for its generalised version (\ref{gen-reina-treves})\footnote{In appendix \ref{app1} the main results of this "procedure" are summarised.}, it is possible to understand which kind of duality transformation of the electromagnetic field is necessary. In fact the actual transformations which produces the NUT generalisation of a given seed spacetime is
\beq
 \label{gen-enhanced-reina-treves}
       \boxed{   \Er \longrightarrow \bar{\Er}_N = \frac{\Er+ib}{1+ib\Er} \qquad \ \ , \ \ \qquad \mathbf{\Phi}\longrightarrow \bar{\mathbf{\Phi}}_N = \frac{\mathbf{\Phi}(1+ib)}{1+ib\Er} } \quad .
\eeq  
Otherwise using the Reina-Treves original notation of the article \cite{reina-treves} the enhanced  transformation reads
\beq
 \label{enanched-reina-treves-trasf}
         \Er \longrightarrow \bar{\Er} = \frac{\xi e^{i\tau}-1}{\xi e^{i\tau}+1}  \quad , \qquad  \mathbf{\Phi}\longrightarrow \bar{\mathbf{\Phi}} = \frac{q \ e^{i\tau}}{\xi e^{i\tau}+1} \quad .
\eeq
The gravitational part of the enhanced Ehlers transformation (\ref{gen-enhanced-reina-treves}) is compatible with the one presented in \cite{nicolai-stelle}, but there electromagnetic fields were not considered. \\ 
Moreover the transformation (\ref{gen-enhanced-reina-treves}) provides better asymptotic properties to the Ernst potentials with respect to the standard Ehlers transformation (\ref{su21-transf})-($III$). In fact, the generic asymptotic expansion of the Ersnt complex potentials for asymptotically flat  metrics (possibly enlarged by the presence of the Taub-NUT parameter) in terms of large radial coordinate $r$, as presented in \cite{alekseev-belinski-kerr}, is
\bea \label{complex-asym} 
       \Er &\sim & 1- \frac{2 \left(M - i B\right)}{r} + \frac{(z_*+2iJ)y+const}{r^2} + O \left(\frac{1}{r^3}\right) \quad ,  \\
    \label{complex-asym2}   \mathbf{\Phi} &\sim &  \frac{Q_e + i Q_m}{r} + \frac{(D_e+iD_m)y+const}{r^2} + O\left(\frac{1}{r^3}\right)     \quad ,
\eea
where $M, B, J, Q_e, Q_m, D_e, D_m,$ respectively identify the total conserved quantities: mass, NUT, angular momentum, total electric and magnetic charge, electric and magnetic dipole moments, while $z_*$ is a constant related to the position of the origin of the quasi-spherical coordinates $(r,\theta)$.\\
It can be easily checked, for instance though the Kerr-Newman-NUT examples of section \ref{Kerr-Newman-NUT-sec}, that the standard Ehlers transformation (\ref{su21-transf}-III) or (\ref{ehlers_Er1})-(\ref{ehlers_Phi1}) is not compatible with the asymptotic fields fall-off at spatial infinity, instead the enhanced Ehlers transformation (\ref{gen-enhanced-reina-treves}), in the context of the Kerr-Newman-NUT solution it is explicitly given in eqs. (\ref{ehlers_ErN}) - (\ref{ehlers_PhiN}), perfectly fulfil the expected decaying of the gravitational and electromagnetic complex potentials. In practice it means that the enhanced Ehlers transformation does not change the values of the other conserved charges, such as electric or magnetic charges, of the given seed spacetime as all the previous variants of the Ehlers transformation do. Thus the enhanced transformation adds the NUT charge while preserving the original physical properties of the initial solution \footnote{Note that mess in the conserved charges introduced cannot be restored by any coordinate transformation or parameter redefinition.}.    
Further evidences of the asymptotic superiority of the enhanced Ehlers transformation are presented in the context of multiple gravitational sources, in section \ref{MPN-section}.\\
The disadvantages of the traditional form of the Ehlers transformation are usually fixed with additional ad-hoc transformations or supplementary boundary conditions. On the contrary, in the examples considered in this paper, both with single and double gravitational sources, the enhanced Ehlers transformation, by construction, does not mix the electric with the magnetic field. \\
Basically, since the enhanced Ehlers transformation, when applied to asymtotically flat spacetimes (as can be also understood from the Kerr-Newman example), preserves the Ernst field fall-off (\ref{complex-asym}) - (\ref{complex-asym2}), it can be considered as the analogous of the electromagnetic duality transformation, but for the gravitational field. In fact in four dimensions the electromagnetic duality rotates the electric monopole charge into the magnetic one, while preserving the spacetime metric and the physical charges such as the mass, the angular momentum or the nut charge\footnote{Very recently, in \cite{huang-kol}, the gravitomagnetic duality is ascribed to be related to a variant of the Janis-Newman trick. Being a complex coordinate transformation, that trick (considered by Ernst a ``method which transcend logic'' \cite{ernst2}) is not a systematic procedure based on the equations of motion, as the solution generating techniques is. Therefore, contrary to the solution generating technique which by construction maps solutions into solutions, when dealing with the Janis-Newman trick, in general, one has not to expect to obtain new metrics or matter fields which fulfil the equations of motion. In the cases the trick may work, for adding the nut charge, it should reduce to the Ehlers transformation.}. Similarly the enhanced Ehlers transformation rotates the mass into the nut charge while preserving the electromagnetic charges; see the appendix \ref{app1} for details in the context of the Kerr-Newman black hole.  \\
Thus the analogy with the electromagnetic duality give us the opportunity to clarify the action of the (enhanced) Ehlers transformation, which effect does not consist merely in adding NUT charge to any axisymmetric and stationary solution. Since the Ehlers transformation rotates the mass charge into the gravomagnetic charge, it means that it is rotating the electric part of the Weyl tensor of the seed metric into it magnetic part. Indeed seed spacetimes with trivial mass charge do not acquire any NUT charge as well, after an Ehlers transformation.  This new insight about the role of the Ehlers transformation naturally opens to explore the possibility of having further dualities stemming from the symmetries of the Ernst equations in general relativity. In fact, at least in restricted setting of the Plebanski-Demianski class of metrics, it is well known that the NUT parameter is the dual of the mass charge, the electric charge is dual to the magnetic monopole and also the rotation and the acceleration parameter are similarly related \cite{plebanski-demianski}. Therefore would be plausible to search for the transformation able to rotate the angular momentum charge into the acceleration parameter, at least in the context of axisymmetric and stationary spacetimes. \\    
To sum up neither the Ehlers nor the Reina-Treves transformations exactly generate the NUT extension of the seed electrovacuum spacetime, but the (\ref{gen-enhanced-reina-treves}) or (\ref{enanched-reina-treves-trasf}) do. Maybe this can be the reason why Reina and Treves in \cite{reina-treves} treated only metrics, such as the Kerr black hole or the Tomimatsu-Sato, with no electromagnetic field. Note also that when trivial and non-trivial transformations are sequentially composed, they may be not reabsorbed by a gauge or a coordinate transformation. For instance the inversion transformation maps the trivial transformation (\ref{su21-transf})-($II$) into the non trivial (\ref{su21-transf})-($III$).  Thus, henceforward in the article, we will prefer to work with the transformation (\ref{gen-enhanced-reina-treves}) to build new solutions, because it is more precise, especially when simultaneously composed with others transformations and because it can be applied to general seeds in the presence of the Maxwell electromagnetic field in a more controlled way, so it is more convenient.\\

\section{NUT generalisation of Majumdar-Papapetrou black holes}
\label{sec-majumbdar}

Now that we have clarified which is the proper transformation able to add the NUT charge to any axisymmetric and stationary electrovacuum space-time, would be interesting to test its effectiveness by  applying the results of the previous section to a given seed to obtain a novel solution.  \\
The generalised enhanced Reina-Treves transformation (\ref{gen-enhanced-reina-treves}) is able to map a static metric in a stationary one. This feature can be exploited to build, for the first time, a coalescence of a binary system composed by two rotating regular black holes at equilibrium. While some binary system describing two Kerr sources at equilibrium have been found \cite{kinnersley}, \cite{dietz}, \cite{neu} \cite{manko-double-kerr-new}, it can be shown that these sources can never describe a legit couple of black holes because one of the two sources have to be hyper-extremal \cite{neugebauer}. If one insists in keeping both  Kerr sources under the extremal limit, non-removable conical singularities between the two black holes, not covered by any horizon, appears. Recently also a couple of regular but accelerating and rotating black holes have been obtained by the Ernst technique \cite{mio-pairs} but the two sources are of a different kind with respect the above examples, as they are not casually connected.\\
Possibly the easiest path to reach our goal is to consider one of the simplest binary regular black hole couple, i.e. the Majumdar-Papapetrou solution \cite{majumdar},\cite{papapetrou} and its non-extremal generalisations. It describes an ensemble of charged and extremal black holes, of the Reissner-Nordstrom type, at equilibrium, where the gravitational attraction is compensated by the electric repulsion between the sources. While the ensemble can be formed by an arbitrary number of sources, for simplicity we will just focus on the easier case composed by two black holes.  \\

\subsection{The non-extremal case: NUT generalisation of a RN black holes pair}
\label{alekseev-beli-nut}

Alekseev and Belinski \cite{alekseev-belinski} and Manko \cite{manko-2007} have been able extend the Majumdar-Papapetrou di-hole solution also beyond the extremal case. But, even in this setting, outside the extremality the two sources, at equilibrium, cannot simultaneously be under-extreme black holes without the introduction of extra conical singularities. Using the notation of  \cite{alekseev-belinski} the double (non-extremal) Reissner-Nordstrom solution can be written in terms of the LWP metric (\ref{lwp-metric}), where the structure functions are
\bea
    \label{f_beli}         f(\rho,z)  &:=&  \frac{\mathcal{D}^2-\mathcal{G}^2+\mathcal{F}^2}{(\mathcal{D}+\mathcal{G})^2}\\
        \omega(\rho,z)  &:=&   0 \\
           e^{2\g(\rho,z)}  &:=&   f_0 \frac{\mathcal{D}^2-\mathcal{G}^2+\mathcal{F}^2}{ \prod_{i=1}^2 (x_i^2-\s_i^2y_i^2)}       
\eea
with $i \in \{1,2\}$ and 
\bea \label{DGF}
     \mathcal{D}  &:=& x_1 x_2 - \bar{\g}^2 y_1 y_2 + \d \left[ x_1^2 + x_2^2 - \s_1^2 y_1^2 - \s_2 y_2^2 + 2 (m_1 m_2 - q_1 q_2) y_1 y_2  \right] \\
      \mathcal{G}  &:=&  m_1 x_2 + m_2 x_1 + \bar{\g} (q_1 y_1 +q_2 y_2) +2 \d \left[ m_1 x_1 +m_2 x_2+ y_1(q_2\bar{\g}-m_1 l) + y_2(q_1\bar{\g+m_2l}) \right]\\
       \mathcal{F}  &:=&  q_1 x_2 + q_2 x_1 + \bar{\g} (m_1 y_1 +m_2 y_2) + 2 \d \left[q_1 x_1 + q_2 x_2+ y_1 ( m_2 \bar{\g} - q_1 l) + y_2(m_1\bar{\g} + q_2 l) \right]
\eea
The bipolar coordinates ($x_i,y_i$) are defined with respect to the coordinates ($\rho,z$) as follows
\bea
            x_i(\r,z) & :=& \frac{1}{2} \left[ \sqrt{\rho^2+(z-z_i+\s_i)^2} + \sqrt{\rho^2+(z-z_i+\s_i)^2}  \right]   \quad ,                    \\
            y_i(\r,z) & :=&  \frac{1}{2\s_i}  \left[ \sqrt{\rho^2+(z-z_i+\s_i)^2} - \sqrt{\rho^2+(z-z_i+\s_i)^2}  \right] \quad .
\eea
The non extremal solution at equilibrium has four independent parameters, the ones related to the masses and charges\footnote{Note that $q_i$ does not coincide with the effective electric charge of the $i$ black hole, which is, according to \cite{alekseev-belinski} $e_1 = q_1- \bar{\g} , \ e_2=q_2+ \bar{\g}$, at least in lack of the NUT parameter.} of the two black holes $m_1,m_2, q_1, q_2$, constrained by the equilibrium condition $m_1m_2=q_1q_2$ and the constant $\bar{\g}$, related to the distance $l$ between the two mass sources placed on the $z$-axis at the points $z_i$
\beq \label{distance-ell}
         l = z_2 -z_1 = (m_2 q_1-m_1q_2)/\bar{\g} \quad .
\eeq
The constants $\s_i$ are connected to the above parameter $\bar{\g}$ by the two constraints $\s_i = m_i^2+\bar{\g}^2-q_i^2$; they determine the position of the horizons, located at $\{ \r=0 , z_i - \s_i \le z \le z_i+\s_i  \}$, which distance is given by $l-\s_1-\s_2$. While the distance between a naked singularity (associated to the source of mass $m_1$) and a black hole (associated to the source of mass $m_2$) is given by $l -\s_2$. Other auxiliary constants are
\beq
           f_0 := \frac{1}{(1+2\d)^2} \quad , \qquad \d := \frac{m_1 m_2 - q_1 q_2}{l^2 - m_1^2- m_2^2 + q_1^2 + q_2^2} \ \ ,
\eeq
but at equilibrium they become trivial: $f_0=1$ and $\d=0$.  
The electromagnetic vector potential supporting this metric is 
\beq \label{A-beli}
            A_\m = \left( \frac{\mathcal{F}}{\mathcal{D+G}} , 0, 0, 0 \right)  . 
\eeq
To obtain the NUT extension of this solution first it is necessary to get its Ernst complex potentials. From definitions (\ref{def-Phi-Er}) we have 
\beq
       \Er = \frac{\mathcal{D-G}}{\mathcal{D+G}}   \qquad , \qquad  \mathbf{\Phi}=  \frac{\mathcal{F}}{\mathcal{D+G}} \quad .
\eeq
Hence applying the enhanced generalised Reina-Treves transformation (\ref{enanched-reina-treves-trasf}) we can add the NUT charge, parametrised by $b$, to the equilibrium configuration of two Reissner-Nordstrom sources. In term of the Ernst potentials it reads
\beq \label{beli-nut}
          \Er_N =\frac{\mathcal{D} (1+ib) - \mathcal{G}(1-ib)}{\mathcal{D} (1+ib) + \mathcal{G}(1-ib)}  \qquad , \qquad \mathbf{\Phi}_N =  \frac{\mathcal{F} (1+ib)}{\mathcal{D} (1+ib)+\mathcal{G}(1-ib)} \quad .
\eeq
As can be easily understood from the non-null imaginary part of the Ernst potential the static Alekseev-Belinski solution after the above transformation becomes stationary. The structure functions for the metric and Maxwell potential can be deduced from eqs. (\ref{def-Phi-Er}) - (\ref{h}), as it was done in the Kerr-Newman-NUT case of section \ref{Kerr-Newman-NUT-sec}. \\
From the asymptotic fall-off of the Ernst fields we can infer the relation between the physical parameters of the seed and of the solution after the Ehlers transformation
\bea
      b   \label{bell2}     & \longrightarrow &   \frac{ - M + \sqrt{M^2+\ell^2}}{\ell}  \quad ,   \\
      m_1  & \longrightarrow & \frac{\left[ q_1 (q_1+q_2) + z_2 (z_2-z_1) \right] \sqrt{M^2+\ell^2}}{(q_1+q_2)^2+(z_1-z_2)^2} \quad ,\\
      m_2 & \longrightarrow &  \frac{\left[ q_2 (q_1+q_2) + z_1 (z_1-z_2) \right] \sqrt{M^2+\ell^2}}{(q_1+q_2)^2+(z_1-z_2)^2} \quad .
\eea
These are the natural multi source generalisation of the parameters rescaling we performed in appendix \ref{app1} to get the Kerr-Newman metric in Boyer-Lindquist coordinates, as it can be seen comparing (\ref{bmell}) with (\ref{bell2}) and (\ref{bmbar}) with 
$$   m_1 + m_2 \longrightarrow \sqrt{M^2+\ell^2}   \quad .    $$
Thanks to the above parameters redefinitions the complex potentials of the double Reissner-Nordstrom-NUT configuration takes the following form at spacial infinity, for $r\approx\sqrt{\r^2+z^2}\rightarrow \infty$
\bea \label{E-2RN}
       \Er &\sim & 1- \frac{2 \left( M - i \ell \right)}{r} + \frac{const}{r^2}  + O\left(\frac{1}{r^3} \right) \quad ,    \\
     \label{Phi-2RN}  \mathbf{\Phi} &\sim &  \frac{2 \left(q_1 + q_2 \right)}{r} +  \frac{(q_1 z_1 + q_2 z_2) [(q_1+q_2)^2+(z_1-z_2)^2- M^2 - \ell^2] y}{[(q_1+q_2)^2+(z_1-z_2)^2] \ r^2} + \frac{const}{r^2} + O\left(\frac{1}{r^3}\right)   \ \ .
\eea

Comparing eqs. (\ref{E-2RN})-(\ref{Phi-2RN}) with the standard asymptotic potential decaying for a generic asymptotically flat solution, as given in (\ref{complex-asym})-(\ref{complex-asym2}) 
we can estimate the values of several physical quantities of the from the first terms of the multipolar expansion of the Ernst field of the double Reissner-Nordstrom-NUT space-time.  In particular the total mass of the gravitational system is given by $M=\sqrt{(m_1+m_2)^2-\ell^2} $, the total electric charge of the space-time is $Q_e= q_1+q_2$, the NUT charge is $\ell$, the magnetic charge $Q_m$ and magnetic dipole moment $D_m$ are null. From the asymptotic fall-off (\ref{E-2RN})-(\ref{Phi-2RN}) it can be read that the generated solution does not acquire any amount of angular momentum after the enhanced Ehlers transformation, i.e. $J=0$, even though it leaves its static state becoming stationary. So, in this case, the stationariness cannot be appreciated from the angular dipole, which being proportional to the angular momentum remains null. But the new parameter introduced by the Ehlers transformation can be appreciated from the contribution to others multipoles, such as the angular monopole or the angular quadrupole. In this sense the resulting spacetime is still twisting.

%To be more precise we present the first four orders of the multipole mass and angular momentum moments, that can be computed, as explained in appendix \ref{app-multipole}
%\bea
 %        M_0 &=&     \nn \\
  %       J_0 &=&
%\eea 
All that is in complete analogy with the well known examples of the Kerr-NUT spacetime, where the angular momentum is not affected by the value of the NUT parameter, which, contributes only from the following orders in the angular multipolar expansion. To clarify this point in the appendix \ref{app1} the firsts angular Geroch-Hansen multipole moments\footnote{These are the same of the Beig-Simon multipole moments, while are equivalent to the Thorne ones only up to a constant factor.} for the Kerr-NUT metric are computed.\\
Finally, from (\ref{complex-asym2}) and (\ref{Phi-2RN}),  we can also deduce the electric dipole of the above solution

$$
    D_e = \frac{(q_1 z_1 + q_2 z_2) [(q_1+q_2)^2+(z_1-z_2)^2- M^2 - \ell^2]}{(q_1+q_2)^2+(z_1-z_2)^2} \quad .
$$

In order to check the regularity of the solution on the axis of symmetry is sufficient to verify wether $\lim_{\r \rightarrow 0} \gamma(\r,z) = 0$. However, as already noted in section \ref{Kerr-Newman-NUT-sec}, the Ehlers transformation applied to the seed metric of the form (\ref{lwp-metric})\footnote{When applied to other variants of the LWP metric, for instance a double Wick rotated version of (\ref{lwp-metric}), which is used, for instance, to obtain black holes in external magnetic field, such as \cite{ernst-magnetic}, the condition on $\g(\r,z)$ is generally not sufficient.} is not affecting $\g(\r,z)$. Therefore the equilibrium condition $m_1 m_2 = q_1 q_2$ might not be modified by the presence of the NUT parameter. Anyway, taking into account the case of the Kerr-Newman black hole, where the NUT charge produces not removable singularities, a more detailed study of this solution is due. In particular, since in this case the metric become stationary too, it is necessary to verify also the regularity on the symmetry axis of the rotational function $\omega(\r,z)$, which turns not-null after applying the Ehlers transformation: $\lim_{\r \rightarrow 0} \omega(\r,z) = const$. However this is outside the scope of the paper, but may be addressed in a future publication; for the rest of the article we will prefer to deal with a simpler case.   \\

\subsection{The extremal case: Majumdar-Papapetrou-NUT spacetime}
\label{MPN-section}

Some stationary generalisations of the Majumdar-Papapetrou solution were firstly discovered by Israel and Wilson \cite{israel-wilson} and indipendently by Perjes \cite{perjes} (IWP), mainly focusing in a multi Kerr-Newman sources\footnote{Also some others multi sources NUT metrics have been presented in \cite{nicolai-stelle}, but without the electromagnetic field, thus these sources cannot sustain some equilibrium configuration}. These proposals fall in a restricted class where the electromagnetic and gravitational Ernst potentials have a very specific linear relation between themself 
\beq \label{wilson-constrain}
        \mathbf{\Phi} = \mathfrak{a} + \mathfrak{b} \ \Er \quad ,  \qquad \mathrm{where} \ \ \ \mathfrak{a} \mathfrak{b}^* + \mathfrak{a}^* \mathfrak{b} = -\frac{1}{2} \quad ,
\eeq 	
for arbitrary complex parameters $ \mathfrak{a} $ and $ \mathfrak{b} $. These assumptions entails a constrained value between the masses and the electric charges of the constituents, which implies extremality only in the absence of the usual angular momentum parameter (often called $a$), otherwise the horizons turn out hyper-extremal. \\
Hawking and Hartle \cite{hartle-hawking} have shown that the only multi sources solutions of these kind, without naked singularities, were the static and extremal ones discovered by Majumdar and Papapetrou. Note that the NUT generalisation of the multi sources non-extremal solution of section \ref{alekseev-beli-nut} does not belong to the family studied by Israel and Wilson. This can be done checking that the Ernst potentials (\ref{beli-nut}) cannot be cast into (\ref{wilson-constrain}); therefore the Hartle and Hawking no-go theorem does not apply for non-extremal metrics.  \\
Hence the extremal case is of special interest between the whole family of two equilibrium Reissner-Nordstrom configurations \cite{alekseev-belinski-4}, \cite{manko-limit-09}, because, only in this specific eventuality, both sources can be considered as regular black holes. For this reason it represents a good choice as a seed solution to analyse our generating technique in more detail. Also the concise and manifest form of the resulting metric, uncommon for stationary multiple gravitational sources, makes apparent the study of some thermodynamic properties for a configuration of double black holes.  \\
For simplicity we choose to deal with an electrically charged seed only (i.e. the standard MP solution). \\
At extremality $m_i=e_i$, $\bar{\g}=\g=0=\s_i$, so the bipolar coordinates simplifies in
\beq
     x_i(\r,z) = \pm \sqrt{\r^2+(z-z_i)^2} \qquad , \qquad y_i(\r,z) = \pm \frac{z-z_i}{\sqrt{\r^2+(z-z_i)^2}} \quad .
\eeq
From eqs (\ref{f_beli})-(\ref{DGF}) it is easy to find that the only non-trivial function of the LWP ansatz becomes
\beq \label{f-beli}
      f(\r,z) = \left(1 + \frac{m_1}{x_1} + \frac{m_2}{x_2} \right)^{-2} = \left[ 1 + \frac{m_1}{\sqrt{\r^2+(z-z_1)^2}} + \frac{m_2}{\sqrt{\r^2+(z-z_2)^2}}\right]^{-2} \quad ,
\eeq
while the electromagnetic potential, after a trivial gauge transformation given by a unitary constant shift with respect to (\ref{A-beli}), can be written as 
\beq
        A_\m = \left[ -  \left(1 + \frac{m_1}{x_1} + \frac{m_2}{x_2} \right), 0 , 0 , 0  \right] \quad .
\eeq
Considering that the two sources are located symmetrically with respect to the origin of the $z$-axis, we can set, without losing generality, that $z_1=-\l$ and $z_2=\l$. In that case the only non-trivial function of the LWP ansatz, describing the Majumdar-Papapetrou solution, takes the economical form
\beq \label{f-manko} 
          f = \l^4\frac{(x^2-y^2)^2}{E_-}= \left[1+ \frac{m_1}{\l} \frac{1}{x+y} + \frac{m_2}{\l} \frac{1}{x-y} \right]^{-2} \quad ,
\eeq
where 
\bea
        E_\pm &=& \l^2 (x^2-y^2) \mp \l x (m_1+m_2) \pm \l y (m_1-m_2) \quad , \\
        F &=& \l \left[ x(m_1+m_2) - y (m_1-m_2)  \right]  \quad ,
\eea 
and the two-dimensional coordinates ($x,y$) can be obtained as the inverse of the usual cylindrical coordinates (of the LWP ansatz) 
\beq \left\{\begin{matrix}
       \r &=& \l \sqrt{x^2-1} \sqrt{1-y^2}     \quad ,  \\
       z &=& \hspace{-2.2cm} \l x y            \quad ,        
\end{matrix}\right.
\eeq
as follows\footnote{More details about this transformation and the differential operators associated to the new set of coordinates can be found in appendix \ref{app2}.}
\beq \label{cor-x-y} \left\{\begin{matrix} 
   x &=& \frac{1}{2\l}  \left[ \sqrt{\rho^2+(z+\l)^2} + \sqrt{\rho^2+(z-\l)^2}  \right] \quad , 
   \vspace{0.15cm} \\
   y &=& \frac{1}{2\l}  \left[ \sqrt{\rho^2+(z+\l)^2} - \sqrt{\rho^2+(z-\l)^2}  \right] \quad .        
\end{matrix}\right.
\eeq
In these coordinates the LWP metric (\ref{lwp-metric}) takes the form
\beq 
  ds^2 = - f(x,y) \big[ dt + \om(x,y) d\varphi \big]^2 + \frac{\l^2}{f(x,y)} \left[  \left(x^2-y^2\right) e^{2\gamma(x,y)} \left( \frac{d x^2}{1-x^2} + \frac{d y^2}{y^2-1} \right) + \left(1-x^2 \right) \left(y^2-1\right) d\varphi^2  \right]   .\nn
\eeq 
Thanks to the coordinates transformation (\ref{cor-x-y}) it is easy to see that the $f$ functions in eqs (\ref{f-beli}) and (\ref{f-manko}) coincide. Moreover the electric field of (\ref{A-beli}) can be cast as $A_t = F/E_-$, while the remaining seed structure functions of the LWP metric are $\omega(x,y)=0$ and $\g(x,y)=0$. Hence the seed Ernst potentials for the Majumdar-Papapetrou solution are
\beq
      \Er_{MP} = \frac{E_+}{F} \qquad \qquad  , \qquad \qquad  \mathbf{\Phi}_{MP} =   A_t = \frac{F}{E_-} \quad .
\eeq
Applying the transformation (\ref{gen-enhanced-reina-treves}) to the above seed we can add to the complex potentials the NUT parameter, as follow
\beq \label{Er-Phi-MPN}
           \Er_{MPN}  = \frac{E_+ + i b F }{ F + i b E_+}   \qquad , \qquad \mathbf{\Phi}_{MPN} = \frac{F^2 (1+ib)}{E_-(F+ibE_+)}   \qquad ,
\eeq 
hence, taking into account the definitions (\ref{def-Phi-Er}), (\ref{A-tilde}) and (\ref{h}), we get\footnote{A Mathematica notebook with the Majumdar-Papapetrou-NUT solution, in several coordinates, can be found at \href{https://sites.google.com/site/marcoastorino/papers/majumdar-papapetrou-nut-black-hole}{https://sites.google.com/site/marcoastorino/papers/majumdar-papapetrou-nut-black-hole}.}
\bea
  \hspace{-0.2cm}    f^{-1} (x,y) &=&   \left[ \frac{m_1}{\l (x+y)} + \frac{m_2}{\l (x-y)} + \frac{1-b^2}{1+b^2}  \right]^2 + \left( \frac{2 b}{1+b^2} \right)^2 \quad , \label{fxy} \\
\hspace{-0.2cm}  \om(x,y) &=&   \frac{4 b}{1+b^2} \left[ \frac{m_2(1-xy)}{x-y} - \frac{m_1(1+xy)}{x+y} \right] \ + \omega_0  \quad ,\\
 \hspace{-0.2cm}       A_t (x,y) &=&   \frac{[m_1(x-y) + m_2(x+y)] \left\{ \frac{1-b^2}{1+b^2} \left[m_1\left(x-y\right) + m_2\left(x+y\right)\right] + \l \left(x^2-y^2 \right)  \right\}}{ \left\{[m_1(x-y) + m_2(x+y)] + \frac{1-b^2}{1+b^2} \l \left(x^2-y^2\right) \right\}^2+ \left[ \frac{2 b \l}{1+b^2} \left(x^2-y^2 \right) \right]^2}   \quad , \label{elecbxy} \\
  \hspace{-0.2cm}    A_\varphi (x,y) &=& \om(x,y) A_t(x,y) + A_{t0}  \label{magnbxy} \quad .
\eea
As explained in section \ref{Kerr-Newman-NUT-sec} the NUT transformation is not changing the decoupled function $\g(x,y)$, which remains null as the seed.\\
It can be easily checked that the Ernst potentials in (\ref{Er-Phi-MPN}) still belongs to the IWP class of solutions, as defined by (\ref{wilson-constrain}), with 
\beq
            \mathfrak{a} = - \mathfrak{b} = \frac{i - b}{2 (i+b)} \quad .
\eeq  
To better understand the physical properties of the solution the polar coordinates centred in one of the two black hole, let's say the black hole of mass $m_1$, 
\bea \label{barr}
     \bar{r} & = &  \sqrt{\r^2+(z+\l)^2} \quad , \\
     \theta  & = &  \arctan \left( \frac{\r}{z+\l} \right) \label{theta-cord} \quad ,
\eea
are useful. $\bar{r} = r - m_1$ and the $\theta$ axis is aligned with the two mass points. In these coordinates the Majumdar-Papapetrou-NUT metric functions $f$ and $\om$ of the LWP metric (\ref{lwp-metric})
\beq \label{lwp-one}
         ds^2 = - f(\bar{r}, \theta) \left[ dt + \omega(\bar{r},\theta)d\phi\right]^2 + f^{-1}(\bar{r},\theta)\left[ d\bar{r}^2 + \bar{r}^2 d\theta^2 + \bar{r}^2 \sin^2 \theta d\varphi^2   \right] \quad ,
\eeq
read respectively
\bea
          f(\bar{r}, \theta)^{-1} &=& \left(\frac{1-b^2}{1+b^2} + \frac{m_1}{\bar{r}}  + \frac{m_2}{\sqrt{\bar{r}^2-4\l \bar{r} \cos \theta +4\l^2}} \right)^2 +  \left(\frac{2b}{1+b^2} \right)^2 \quad ,\label{f-one} \\
          \omega (\bar{r}, \theta) &=&  \frac{4b}{1+b^2} \left[ - m_1 \cos \theta + m_2 \left( \frac{2\l - \bar{r} \cos \theta }{\sqrt{\bar{r}^2-4\l \bar{r} \cos \theta +4\l^2}} \right) \right] \quad . \label{omega-one}
\eea
While also the associated electromagnetic field in coordinates $(\bar{r},\theta)$ can be easily found from eqs (\ref{elecbxy}) and (\ref{magnbxy}) by a coordinates transformation that can be obtained by combining the two transformations (\ref{cor-x-y}) and (\ref{barr})-(\ref{theta-cord}), as shown in appendix \ref{app2}. In these coordinates is very clear that the physical boundary conditions for the asymptotic decaying of the Ernst potential additionally imposed by Israel and Wilson, 
\beq
         \Er   \xrightarrow[\bar{r} \to \infty]{} 1 \quad \quad , \qquad  \qquad \ \mathbf{\Phi}  \xrightarrow[\bar{r} \to \infty]{} 0  \qquad ,
\eeq
are automatically implemented by the enhanced Ehlers transformation. On the other hand this is not true for the standard Ehlers transformation, where the gravitational Ernst potential $  \Er $ goes asymptotically to a complex constant, when applied to the Majumdar-Papapetrou solution.  \\ 
The limits to the well known solution, such as the extremal Reissner-Nordstrom or the static Majumdar-Papepetrou are very clear. To get the extremal Reissner-Nordstrom it is sufficient to vanish the second black hole, through the $m_2 \rightarrow 0$ limit, while the Majumdar-Papepetrou, in the standard coordinates presented in \cite{hartle-hawking}, \cite{chandra} is retrieved when the nut parameter $b$ is null (the relation between the nut parameter $b$ of the enhanced and generalised Reina-Treves transformation is related to the standard NUT parameter as seen in eq (\ref{bmell})). \\
The two event horizons are located at $\r=0$ and $z=\pm \l$, thus their distance is $2\l$. Actually, in terms of the radial coordinate centred in the $ m_1 $ black hole, the loci identified by $ \r = 0 $ and $ z = - \l $ correspond to $ \bar{r} = 0 $, which is not a point but it represents a surface of finite area 
$$ \mathcal{A}_1 \ \bigg|_{\bar{r}=0} = \int_0^{2\pi} d\varphi\int_0^{\pi} d\theta \sqrt{g_{\theta \theta} g_{\varphi\varphi}}  \ \bigg|_{\bar{r}=0} =  \ 4 \pi m_1^2 \quad . $$
Moreover from the inspection of the polar circumference (meridian)
\beq
            C_p =\int_0^{2\pi} \sqrt{g_{\theta \theta}} d\theta \ \bigg|_{\bar{r}=0} = 2 \pi m_1
\eeq
and azimuthal circumference (parallels)
\beq
            C_a =\int_0^{2\pi} \sqrt{g_{\varphi\varphi}} d\varphi\ \bigg|_{\bar{r}=0} = 2 \pi m_1 \sin \theta
\eeq
of the event horizon of mass $m_1$ (at constant $t$) it is possible to infer the spherical symmetry of the event horizon.\\
Specular result for the event horizon area and circumferences can be obtained using a set of coordinates centred in the second black hole, just replace $m_1$ with $m_2$. \\
Following the reasoning of Hartle and Hawking in \cite{hartle-hawking} for the NUT free case, it is possible to show that the geometry at $\bar{r}=0$ is regular and it describes a null surface. At this scope consider the following coordinates transformation
\beq
            t = u + F(\bar{r})  \quad ,
\eeq
where the $F(\bar{r})$ function is chosen such that 
\beq
           \frac{d F}{d \bar{r}} = \left[ \sqrt{f^{-1}(\bar{r},\theta)} \Big|_{r\approx 0} \right]^2 =\left[ \frac{m_1}{\bar{r}} + \frac{m_2 (1+b^2) + 2 \l (1-b^2)}{2\l(1+b^2)} \right]^2 =: V^2(\bar{r})\quad .
\eeq
Then, defining $U(r):= \sqrt{f^{-1}(\bar{r},\theta)}$, the metric becomes
\bea
        ds^2 &=& -\frac{du^2}{U^2(\bar{r})} + d\bar{r}^2 \left[ \frac{U^4(\bar{r}) - V^4(\bar{r})}{U^2(\bar{r})} \right] - 2 \frac{V^2(\bar{r})}{U^2(\bar{r})} du d\bar{r} + 2 \frac{\omega(\bar{r}, \theta)}{U^2(\bar{r})} du d\varphi + 2 \frac{V^2(\bar{r})}{U^2(\bar{r})} d \bar{r} d\varphi  \nn \\ 
            & & + \frac{\omega^2(\bar{r}, \theta)}{U^2(\bar{r})} d\varphi^2 + U^2(\bar{r}) \ \bar{r}^2 \left( d\theta^2 + \sin^2 \theta \  d\varphi^2 \right) \quad ,
\eea
which, in the neighbourhood of $\bar{r}=0$, it simplifies to
\bea
       ds^2 &=& - \frac{\bar{r}^2}{m_1^2}  du^2 + \left[ \frac{8 b^2 \l^2 + (1+b^2)^2 m_1 m_2 \cos \theta}{1+b^2} \right] d\bar{r}^2 - 2 du d\bar{r}  \\
       & & + \left[ \frac{m_2 + m_1 \cos \theta}{1+b^2} \right] 8b \left[ \frac{\bar{r}^2}{m_1^2} du d\varphi + d\bar{r} d\varphi + 2b \frac{\bar{r}^2}{m_1^2} d\varphi^2 \right] + m_1^2 \left( d\theta^2 + \sin^2 \theta d\varphi^2 \right) \quad . \nn 
\eea
The above metric is explicitly regular at $ \bar{r}=0 $, thus we can extend the manifold to negative values of $\bar{r}$, furthermore it's clear that $\bar{r}=0$ describes a null-surface. Thanks to the following definitions
\beq
              r_1 = - \bar{r} \qquad , \qquad r_2 = \sqrt{r_1^2 + 4 \l r_1 \cos \theta +4\l^2} \quad 
\eeq  
the extended spacetime is described by the metric (\ref{lwp-one}) but with $f(\bar{r},\theta)$ replaced by 
\beq \label{ftildameno1}
      \tilde{f}(r_1,\theta)^{-1} := \left(\frac{1-b^2}{1+b^2} - \frac{m_1}{r_1}  + \frac{m_2}{r_2} \right)^2 +  \left(\frac{2b}{1+b^2} \right)^2 \quad .
\eeq
In the known case of standard Majumdar-Papapetrou metric, for $b=0$, let's call the content of the non-vanishing bracket of (\ref{ftildameno1}) $\tilde{U}_0(r_1,\theta)$. In that case the function $\tilde{U}_0(r_1,\theta)$, close to $r_1=0$ takes large negative values, while for $r_2 \approx 0$ takes positive values, so it must vanish somewhere in between, where the metric diverges. 
Actually  from the curvature scalar invariants\footnote{The curvature scalar invariants are computationally easier to obtain in the $(x,y)$ coordinates.}, it is possible to realise that the zeros of $\tilde{U}_0$ are points where the curvature invariants diverge, while as we have seen above, $r_1=0$ it is just a coordinate singularity.
%For large values of $r_1$ and $r_2$ the function $\tilde{f}^{-1}(r_1,\theta)\approx 1$, so the surface   
The locus of $\tilde{U}_0(r_1,\theta)=0$ is not a surface but a point. It can be shown, just considering a surface inside the domain defined by $\tilde{U}_0(r_1,\theta)=0$ and letting it tend to $\tilde{U}_0(r_1,\theta)=0$, we get
\beq
       \int_0^{2\pi} d\varphi\int_0^{\pi} d\theta \  \tilde{U}^2_0(r_1,\theta) r_1^2 \sin \theta \quad   \longrightarrow \ \ 0    \quad .
\eeq
So, for $b=0$, the region interior to $r_1=0$ shows similarities with the interior of the RN black hole. \\
When $b \ne 0 $ the function $\tilde{f}^{-1}(r_1,\theta)$ cannot vanish therefore the metric, thanks to the nut parameter $b$ becomes regular, exactly as occurs in the case of a single NUT black hole, where the curvature singularity is smeared by the NUT parameter. Therefore the radial coordinate $\bar{r}$ can be continued without obstructions for negative values. This feature recently makes people think that black holes generalisation with the NUT parameter are suitable, under certain assumptions, to traversable wormhole interpretation\footnote{While some constraints on the upper bound of the NUT parameter were proposed, the presence of closed-timelike curves, that violate causality, for free-falling observers cannot completerly discarted in \cite{clement-nut-wormhole}.} \cite{clement-nut-wormhole}. %Actually in \cite{clement-multi} some axisymmetric solutions representing a collection of  wormhole with NUT charge were presented. \\
\\

\subsection{Warm up: Reissner-Nordstrom-NUT/CFT correspondence}
\label{app-RN-NH}

The Kerr-CFT correspondence aims to describe the microscopical degrees of freedom of a black hole, especially at extremality, through a duality between the near horizon geometry of the black hole and a conformal field theory  dual to the near horizon geometry placed on its asymptotic region,\cite{strom-review} and \cite{compere-review}.   \\
It would be interesting to check if the tools provided by the Kerr-CFT can be extended in the context of multi black holes.\\
From the extremal near horizon geometry of the Majumdar-Papapetrou-NUT black hole it is possible to  extract the central charge of the dual CFT associated to the black hole solution. Unfortunately in the case of extremal charged black hole with NUT the diagonal form of the near horizon metric makes difficult to implement the standard stationary Kerr-CFT picture, where the central charge is strongly related to the rotation of the black hole.\\
On the other hand, for the electrically charged black hole solutions, it is possible to exploit an alternative approach which is not based on the rotational symmetry around the azimuthal angle, but on the $U(1)$ symmetry of the Maxwell gauge potential. The Kaluza-Klein uplift of our solution in one dimension more transforms the electromagnetic degrees of freedom in rotational degrees of freedom, as explained in \cite{compere-review} (see also \cite{magn-RN-CFT} and \cite{marcoa-acc}). In fact, considering the Abelian gauge field to be wrapped around a compact extra dimension $\psi$ of period $2 \pi R_\psi$, we can define an extremal Frolov-Thorne vacuum, which corresponds to a temperature in the boundary conformal field theory\footnote{Since on the dual CFT model the Frolov-Thorne vacuum plays the role of a temperature it is often called Frolov-Thorne temperature.}, in units of $R_\psi$: $T_\psi= T_e R_\psi$. In analogy with the standard rotational picture, at extremality, the electromagnetic chemical potential $T_e$ can be defined as 
\beq \label{Te}
         T_e := \lim_{r_+ \rightarrow r_+^{ext}} \frac{T_H}{\Psi_e^{ext}-\Psi_e} \quad,
\eeq
where the $\Psi_e$ represents the Coulomb electromagnetic potential evaluated at the event horizon ($\Psi_e^{ext}$ is the extremal case), while $T_H$ is the usual Hawking temperature.\\
To clarify some points let's review some aspects of the microscopic entropy the Reissner-Nordstrom-NUT black holes, recently studied also in \cite{sakti}. The line element and the gauge vector potential of the dyonic generalisation of the solution found by Brill in \cite{brill} are given respectively by 
\beq
      ds^2 = - \left( \frac{r^2-2mr+q^2+p^2-\ell^2}{r^2+\ell^2} \right) \left[ dt+ 2 \ell \cos \theta d\varphi \right]^2 + \frac{dr^2}{\frac{r^2-2mr+q^2+p^2-\ell^2}{r^2+\ell^2} }
+ (r^2+\ell^2) \left[ d\theta^2 + \sin^2 \theta \ d\varphi^2 \right] \quad \eeq
and
\beq
      A_\m = \left[ - \frac{q r + p \ell}{r^2+\ell^2} , 0 , 0 ,  \frac{ p ( r^2 - \ell^2) -2 q r \ell}{r^2+\ell^2} \cos \theta \right] \quad .
\eeq
The metric presents two Killing horizons, an inner $r_-$ and outer $r_+$ one, located at $r_\pm=m \pm \sqrt{m^2-q^2-p^2+\ell^2}$. In the extremal case they coincide at $r_+^{ext} = \sqrt{q^2+p^2-\ell^2}$.\\
Since the angular velocity at the event horizon is null, $\Omega_J=0$, the Killing vector generating the event horizon remains the same of the Reissner-Nordstrom one: $ \chi = \p_t $.
Then the Hawking temperature $T_H$ is defined from the surface gravity $\k_s$ as follows
\beq 
       T_H  = \frac{\k_s}{2\pi} = \frac{1}{2 \pi} \sqrt{-\mezzo \nabla_\m \chi_\n \nabla^\m \chi^\n } \ \Bigg|_{r_+} = \frac{1}{4\pi} \frac{r_+-r_-}{r^2_+ + \ell^2} \quad .
\eeq
Thus at extremality $(r_+=r_-)$ the Hawking temperature become null. Nevertheless the electromagnetic chemical potential, as defined in (\ref{Te}), is well defined and non null for the extreme RN-NUT black hole
\beq
      ^{[RNN]}T_e = \frac{1}{2\pi} \frac{p^2+q^2}{p^2 q+q^3-2 q \ell ^2+ 2 p \ell  \sqrt{p^2+q^2-\ell ^2}} \quad ,	
\eeq
since the Coulomb electromagnetic potential for RN spacetime is given by
\beq 
         ^{[RNN]}\Psi_e = -\chi^\m A_\m \ \Bigg|_{r_+} = \frac{qr_++p\ell}{r_+^2+\ell^2} \quad .
\eeq
To reach the extremal RN-NUT near horizon metric the following change of coordinates adapted to the horizon is needed
 \beq \label{near-hor-coord-transf}
               r(R):= r_+^{ext} + \b r_0 R \quad , \quad t(\tau):= \frac{r_0}{\b} \tau \quad , \quad \varphi(\tau, \phi) := \phi+ \Omega_J \frac{r_0}{\b} \tau  \quad .
 \eeq
Taking the limit for $\b \rightarrow 0$ the metric, near the horizon of an extreme RN-NUT black hole, takes the form of warped product of $AdS_2\times S^2$. In particular, as shown in \cite{lucietti-reall} or \cite{compere-review}, it can be cast into the standard form of an extremal near-horizon geometry    
\beq
       ds^2 = \G(\theta) \left[ -R^2 d\tau^2 + \frac{dR^2}{R^2} + \alpha^2(\theta) d^2\theta + \gamma^2(\theta) \big( d\phi + \k R d\tau  \Big)^2 \right] \quad ,
\eeq 
for 
$$\G(\theta)=q^2+p^2 \quad , \qquad  \a(\theta)=1 \quad , \qquad \k=0 \quad , \qquad \g(\theta)=\sin \theta \quad .$$ \\
To write the near horizon gauge vector potential, in addition to the coordinates transformation (\ref{near-hor-coord-transf}), the electric potential have to be gauge shifted: $A_t \longrightarrow A_t+\Psi_e$. Then in the near-horizon limit, for $\b \rightarrow 0$ we have
\beq
       A_\m = \left[R \ \frac{p^2 q+q^3-2 q \ell ^2+ 2 p \ell  \sqrt{p^2+q^2-\ell ^2}}{p^2+q^2}  , 0 , 0 , \cos \theta \ \frac{q^2 p + p^3 - 2 p \ell ^2 - 2 q \ell  \sqrt{p^2+q^2-\ell ^2}}{p^2+q^2} \right] \ \ .
\eeq 
Comparing the above potential with the general near horizon form for the static case,
\beq
          A = \bar{e} R d\tau - \frac{\bar{p}}{4\pi} \cos \theta d\phi \quad ,
\eeq
we can extract the value of the constant 
\beq
               \bar{e} = \frac{p^2 q+q^3-2 q \ell ^2+ 2 p \ell  \sqrt{p^2+q^2-\ell ^2}}{p^2+q^2} \quad .
\eeq
Hence the electromagnetic chemical potential for the RN-NUT black hole can be written as
\beq  \label{Te=1/e}
                 ^{[RNN]}T_e = \frac{1}{2\pi \bar{e}}  \quad .
\eeq
Finally the central charge of the dual conformal field model living on the boundary of the near horizon metric, as explained in \cite{compere-review}, \cite{marcoa-acc} is given by
\beq
            c_Q = \frac{3 \bar{e}}{R_\psi} \int_0^{\pi} \G(\theta) \g(\theta) \a(\theta) d\theta =  \frac{6 \bar{e}}{R_\psi} (q^2+p^2) \quad .
\eeq 
According to the Kerr/CFT correspondence, the microscopic entropy of the extremal RN-NUT black hole can be obtained by the Cardy formula, considering as the left central charge the $c_Q$ and as the left temperature\footnote{The right temperature is null for extreme configuration.} $T_\psi$ 
\beq
       \mathcal{S}_{CFT} = \frac{\pi^2}{3} c_L T_L  = \pi (q^2+p^2) = \frac{\mathcal{A}}{4} \quad .
\eeq
So we get that the microscopic entropy of the extremal Reissner-Nordstrom-NUT coincides with a quarter of its horizon area $\mathcal{A}$
\beq
         \mathcal{A} = \int_0^{2\pi} d \varphi \int_0^{\pi} d\theta \sqrt{g_{\theta \theta} g_{\varphi \varphi} } = 4 \pi (r_+^2 + \ell^2) \Bigg|_{r_+\rightarrow r_+^{ext}} = 4 \pi (q^2+p^2) \quad .
\eeq
Note that, even though the presence of the NUT parameter modifies the position of the horizons, it is not affecting the final result: the entropy remain the same of the standard dyonic extreme Reissner-Nordstrom black hole. Note also that the microscopic entropy of RNN black hole, according to the Kerr/CFT correspondence agrees with the Bekenstein-Hawking law, but it does not agree with others entropy proposals, such as  \cite{Garfinkle-Mann}.\\

\subsection{Near horizon geometry of the Majumdar-Papapetrou-(NUT) spacetime}

As we have seen in section \ref{app-RN-NH}, for a single extremal charged black hole it is known that the near-horizon geometry is not affected directly by the presence of the NUT parameter. It would be interesting to see if the same result holds also for the double extremal charged black hole endowed with the NUT charge. At this scope let's consider the di-hole solution in the coordinates centred in one of the black holes, given by the eqs (\ref{lwp-one})-(\ref{omega-one}). To get the near horizon limit it is propaedeutic a change of coordinates adapted (and in a co-rotating frame with respect) to the event horizon
\beq
       \bar{r}(R):= \bar{r}_+ + \b r_0 R \quad , \quad t(\tau):= \frac{r_0}{\b} \tau \quad , \quad \varphi(\tau, \phi) := \phi+ \Omega_J \frac{r_0}{\b} \tau  \quad .
\eeq
The constant $r_0$ is needed to remain with dimensionless coordinates, while $\Omega_J$ is the angular velocity at the horizon $r_+$, which results null even though the metric is stationary:
\beq 
        \Omega_J := - \frac{g_{\t\phi}}{g_{\phi\phi}} \Bigg|_{r_{+}} = 0 \quad .
\eeq
As explained above, in these coordinates the event horizon is where $\bar{r}$ vanishes, i.e. $\bar{r}_+ = 0 $. Taking the limit for $\b \rightarrow 0$, and choosing $r_0=m_1$, we obtain the geometry of the Majumdar-Papapetrou-NUT spacetime near the horizon of the black hole of mass $m_1$
\beq \label{nh-m1}
      \tilde{ds}^2 = m_1  \left[ - R^2 d\tau^2 + \frac{dR^2}{R^2}  + d\t^2 + \sin^2 \t d\phi^2   \right]  \quad .
\eeq
It's worth to point out that this near horizon geometry coincides exactly with the one of the extreme Reissner-Nordstrom-(NUT) black hole and that it does not depends on the NUT parameter, as in the single black hole case. Also note that in the presence of an ensemble of black holes, the geometry near the event horizon of a single black hole still falls in the standard class of the extremal near horizon geometries described in \cite{lucietti-reall}. In the case under consideration, we can infer, from eq. (\ref{nh-m1}), that the near horizon metric represents a warped product of $AdS_2 \times S^2$, thus it is endowed with the $SL(2,\mathbb{R}) \times U(1)$ isometry. \\
Of course a similar result can be obtained using coordinates centred in the black hole of mass $m_2$, just replacing $m_1$ with $m_2$. Actually from the Majumdar-Papapetrou-NUT solution written in coordinates $(\r,z)$,\footnote{See appendix \ref{app-rho-z} for details.} it is possible to simultaneously write both near horizon geometries, when the mass of the two black hole coincides, i.e. $m_2=m_1$, considering the near-horizon change of coordinates
\beq
                      \r(R,\theta) := \b r_{0_\mp} R \sin \t \quad , \qquad z(R,\t):= \mp \l + \b r_{0_\mp} R \cos \t \quad , \qquad t(\tau):= \frac{r_{0_\pm}}{\b} \tau \quad ,
\eeq
and performing the $ \b \rightarrow 0 $ limit. The resulting near horizon geometry is described by the metric 

\beq \label{nh-m1m2}
      \tilde{ds}^2 = m^2_i  \left[ - R^2 d\tau^2 + \frac{dR^2}{R^2}  + d\t^2 + \sin^2 \t d\phi^2   \right]  \quad ,
\eeq

where $i=1$ is associated to the source centred in $(\r=0,z=-\lambda)$, while $i=2$ to the one localised by $(\r=0,z=+\lambda)$. When the masses of the sources coincide ($m_1=m_2$), then $r_{0_{-}}=r_{0_{+}}=r_0=m_1$, when the masses do not coincide  $r_{0_{-}}=m_1$ and $r_{0_{+}}=m_2$.  As always occurs for regular extremal black hole this metric models an $AdS \times S^2$ spacetime.\\
To obtain the near horizon form of the electromagnetic potential we firstly perform the usual gauge transformation on the vector potential $A_t \rightarrow A_t + \Psi_e$ of the non-extremal solution (\ref{A-beli}); then in the coordinates adapted to the horizon we take the $\b \rightarrow 0$ limit, which gives
\beq
             A = \bar{e} R d\tau   \qquad , \qquad \textrm{where} \qquad \bar{e} = \frac{r^2_{0_\mp}} {m_{\substack{1\\ 2}} }\quad .
\eeq
Following the methods provided by the Kerr/CFT framework \cite{strom-review}, \cite{compere-review} it is possible to evaluate the, so called, microscopic entropy of the Majumdar-Papapetrou di-hole. In the case under consideration multiple disconnected event horizons are present therefore the procedure, originally found for a single black hole, have to be somehow adapted. The more straightful case is certainly when the two extremal sources have the same mass, because in that case the boundary condition of the near horizon geometry can be considered on the same footing. On the other hand, when the two sources are not identical, the easier approach is to treat each one individually. According to the standard prescription 
given by the Kerr/CFT correspondence, the microscopic entropy of the extremal black hole is associated to the (left) temperature and the central charge of the associated modular invariant conformal field theory model located on the boundary of the near horizon solution, can be derived by the Cardy formula\footnote{For the details concerning the applicability of the Cardy formula we refer to the review \cite{compere-review}.}
\beq \label{cardy}
          \mathcal{S}_{CFT} = \frac{\pi^2}{3} c_L T_L  \quad .
\eeq
As we have seen in the single black hole case studied in section \ref{app-RN-NH}, the left temperature of the boundary conformal field theory is associated with the chemical potential $T_e$, defined in (\ref{Te}) and  (\ref{Te=1/e}).
Because the Majumdar-Papapetrou di-hole is associated to a couple of conformal systems we treat the above quantities additively, as follows
\bea
        c_L &=& \sum_{i=1}^2 \ \frac{3 \bar{e}_i}{R_\psi} \int_0^{\pi} \G_i(\theta) \g(\theta) \a(\theta) d\theta \ = \ \frac{6 \bar{e}_i}{R_\psi} m_i^2 \quad , \\
        T_L &=& \sum_{i=1}^2 \ R_\psi T_{e_i} = \sum_{i=1}^2 \ \frac{R_\psi}{2 \pi \bar{e}_i}  \quad . \label{TL}
\eea
In this way, from (\ref{cardy})-(\ref{TL}), the resulting entropy of the conformal field model associated to the double black hole ensemble described by Majumdar-Papapetrou solution is given by
\beq
     \mathcal{S}_{CFT} = \sum_{i=1}^2  \pi m^2_i = \pi (m^2_1+m^2_2) = \frac{\mathcal{A}_1+\mathcal{A}_2}{4}  = \frac{\mathcal{A}}{4} \quad .
\eeq  
As expected, the entropy of the gravitational system, inferred by the conformal field techniques and the near horizon analysis, coincides with a quarter of the total area $\mathcal{A}$ of the event horizons of the two black holes, in agreement with the Bakenstein-Hawking formula.\\
Lastly we stress the utility of the non-extremal solution of section \ref{alekseev-beli-nut}, in the above framework, because the chemical potential $T_e$ is defined through a limit from non-extremal quantities. \\

\subsection{Second Law of Thermodynamics}

Even though the solution we generated is not dynamical and therefore it can not describe the collision and merging process of the two sources, as time changes, we still can estimate which of the two configurations described either by the disjoint ensemble of two black holes of Majumdar-Papapetrou type or by the collapsed configuration, is thermodynamically favoured; specifically which of the two has the bigger entropy.\\
The collapsed configuration can be obtained just by letting the distance between the two sources, $l$, as defined in (\ref{distance-ell}), going to zero, i.e. $\l \rightarrow 0$. In that case the resulting black hole has the total mass and electric charge given by the sum of the ensemble components. Note that both configurations have the same total mass and total electric charge.\\ 
Extremal black hole entropy is still a puzzling issue even for a single source. In fact, while from the semiclassical action approach \cite{hawking-ross}, \cite{teitelboim}, \cite{gibbons-kallosh} the extremal black hole entropy seems to be null, from a microscopical point of view, according to string theory \cite{strominger-vafa} or the Kerr/CFT correspondence \cite{strom-review}, the black hole entropy follows the usual Bekenstein-Hawking formula, i.e. a quarter of the event horizon area. Works in the licterature trying to clarify the discrepancy, such as \cite{mitra} and \cite{carroll}, point to  validate the Bekentein-Hawking area law also at extremality.    \\
Black hole entropy becomes even a more puzzling issue when the NUT parameter deforms the extremal black hole, because there are different results, where the NUT parameter plays an active role or not, depending also on the interpretation of the Misner string related to these spacetimes, \cite{mann-recent}. \\
In section \ref{app-RN-NH} it is shown how, according to the Kerr/CFT correspondence approach, for a single extremal and charged black hole with NUT the Bekenstein-Hawking formula holds (without any role played by the NUT parameter). Furthermore we have seen, in the previous subsection, as the same computation for the entropy of the double source solution can be also realised.  \\
As the strong similitude between the near horizon geometry of the single and of the double extremal systems suggests,  we have found that also the entropy of the Majumdar-Papapetrou-(NUT) black holes is given by a quarter of its event horizon area. \\
Actually even without relying on the duality relation with the conformal field theory, just from the near-horizon metric $\tilde{ds}$, as in (\ref{nh-m1m2}), it is very easy to deduce the areas $\mathcal{A}_1$ and $\mathcal{A}_2$ of each of the couple of black holes, which is 
\beq
        \mathcal{A}_i  = \int_0^{2\pi} d\phi\int_0^{\pi} d\theta \sqrt{\tilde{g}_{\theta \theta} \tilde{g}_{\phi\phi}}   =  \ 4 \pi m_i^2 \quad .
\eeq
Hence, assuming the validity of the Bekenstein-Hawking entropy (as shown above in the framework of the Kerr/CFT correspondence), the total entropy provided by the system of two separated extremal black holes $\mathcal{S}_{\odot\odot}$ come out to be proportional to the sum of the two horizon surfaces  
\beq
         \mathcal{S}_{\odot \odot} = \frac{1}{4} \ \sum_{i=1}^2 \mathcal{A}_i %= \frac{\mathcal{A}_1}{4} + \frac{\mathcal{A}_2}{4} 
         = \pi (m_1^2+m_2^2) \quad .
\eeq
On the other hand the entropy $\mathcal{S}_{\bigodot}$of the system where the two black holes are joint is given by
\beq
            \mathcal{S}_{\bigodot} = \frac{\mathcal{A}_{\bigodot}}{4} = \pi (m_1+m_2)^2 \quad .
\eeq
Therefore, for the same values mass and electric charge, the collapsed single black hole is the thermodynamically favoured configuration, due to its higher entropy 
\beq  \label{2-law}
       \mathcal{S}_{\odot\odot}  \ \leq  \  \mathcal{S}_{\bigodot}     \quad , 
\eeq
in agreement with the Hawking area law, which states that from a classical point of view\footnote{With classical we mean we are ignoring possible quantum mechanical effects that might cause the evaporation of black hole, thought emission of Hawking radiation.} the even horizon area of a black hole tends to increase.
Hence this relatively simple model allows us to confirm that the second law of black hole thermodynamics is verified for extremal (and charged) binary merging. A similar result have been found recently for a different couple of black holes held at equilibrium by a conical singularity \cite{maria}.\\
Note that the equality in eq (\ref{2-law}) holds only if one of the two black holes disappears, which occurs when its mass parameter vanish; for all other proper double source configuration only the strict inequality holds. \\ 
In all this thermodynamics discussion the presence of the NUT parameter is quite irrelevant, as it does not directly plays a role in the final values of the charged black hole entropy (nor area), as it occurs also in the single source seen in section \ref{app-RN-NH}. Therefore we can state that, also for the standard double Majudar-Papapetrou di-hole spacetime, the more entropic state is given by the collapsed one.  \\

\section{Summary, Comments and Conclusions}

In this article a detailed study of the Ehlers transformation is presented. It is shown, through explicit examples,  why the known methods used to add the NUT parameter to an axisymmetric and stationary spacetime in general relativity are not precise when the Maxwell electromagnetic field is coupled to the gravitational theory. \\
We have shown how to modify the Ehlers and the Reina-Treves transformations in order to generalise an electrovacuum spacetime by adding the NUT parameter. The transformation is also able to remove the NUT charge when it is applied to a seed endowed with an undesired NUT parameter, without deforming the electromagnetic potential. \\
By analysing the Ernst potentials generated by the new transformation proposed we understand that it acts as a duality transformation between the mass and the gravitomagnetic charge, in a similar way the duality transformation for the Maxwell electromagnetism rotates the electromagnetic field. As a future perspective would be interesting to deepen this point and, in particular, to explore the possibility of finding other kinds of dualities for the gravitational theory, even outside of the axisymmetric and stationary setting. \\  
We have applied this enhanced transformation of the Ernst equations to obtain a new solution representing a NUT generalisation of a couple of charged black holes at equilibrium. The extremal limit of this metric gives the Majumdar-Papapetrou-NUT spacetime.\\
Moreover thanks to the near horizon techniques some thermodynamic properties of the Majumdar-Papapetrou-(NUT) solution have been studied. Since the solution, when varying the parameters, can describe both a single or a double black hole event horizon, it is possible to confirm which configuration of the system is favoured from a thermodynamic point of view. The outcome is in agreement with the second law of black hole thermodynamics because, for a given mass, the bigger entropy configuration is obtained when the parameters of the solution are such that the two black hole are collapsed in a single one, which have a bigger area as well.\\  
The extremal character of the Majumdar-Papapetrou-(NUT) solution provide a good testing ground to prove the applicability of the techniques borrowed from the Kerr/CFT correspondence also for multi black hole configurations. In particular it was possible to reproduce the microscopic entropy for the extremal charged di-hole system. \\  
We observe that, following \cite{embedding-marcoa} - \cite{stationary-marcoa}, the procedure illustrate here is easily generalisable to the presence of a minimally or a conformally coupled scalar field\footnote{In particular the NUT generalisation of the Bekenstein black hole is worked out in \cite{charmo}.} (and all the related scalar-tensor gravitational theories connected with them by a conformal transformation, such as some Branse-Dicke or $f(R)$ gravities).  
On the other hand the generalisation to the presence of the cosmological constant is not that easy because the alleged breaking of the symmetries of the Ernst equations \cite{marcoa-lambda}. \\

\section*{Acknowledgements}
{\small I would like to thank Vladimir Manko for fruitful discussions and for suggesting references. \\
This work has been funded by Fondecyt project n$^\textrm{o}$ 11160945 and partially by the Conicyt — PAI grant n◦79150061, by Beca Chile, INFN and by MIUR-PRIN contract 2017CC72MK-003.}\\

\appendix

\section{Notation: Differential operators in various coordinates}
\label{app2}

Let's fix the notation of section \ref{Kerr-Newman-NUT-sec}, in particular the definition of differential operators in various coordinate systems. One of the advantages of writing the Einstein-Maxwell equations using the Lewis-Wayl-Papapetrou ansatz in coordinates $(\r,z)$ is that the gravitational equations, which usually are written by curved differential operators, can be easier written in terms of flat differential operators. The same non-trivial property is inherited also by the complex Ernst equations.\\
For this reason we are interested in flat three-dimensional spacetime in cylindrical coordinates $(\rho,z,\varphi)$, whose metric can be written as 
\beq
         ds^2 = d\r^2 + dz^2 + \r^2 d\varphi^2 \quad .
\eeq
For any scalar $f(\r,z,\varphi)$ or vectorial function $\overrightarrow{A} (\r,z,\varphi)$, the gradient, the divergence  and the laplacian are respectively
\bea \label{cily-diff1}
      \overrightarrow{\nabla} f(\r,z,\varphi) &=& \overrightarrow{e}_\r \frac{\p f(\r,z,\varphi) }{\p \r} + \overrightarrow{e}_z \frac{\p f(\r,z,\varphi) }{\p z} + \overrightarrow{e}_\varphi \frac{\p f(\r,z,\varphi) }{\p \varphi}   \quad ,  \\
       \overrightarrow{\nabla} \cdot \overrightarrow{A}(\r,z,\varphi) &=&  \frac{1}{\r} \frac{\p}{\p \r} \big( \r A_\r \big) + \frac{\p A_z}{\p z} + \frac{\p A_\varphi}{\p \varphi}  \quad , \\
     %  \overrightarrow{\nabla} \times \overrightarrow{A}(\r,z,\varphi) &=&      \\
     \label{cily-diff3} \nabla^2 f(\r,z,\varphi) &=&  \frac{\p^2 f}{\p \r^2} + \frac{\p^2 f}{\p z^2} + \frac{1}{\r^2}  \frac{\p^2 f}{\p \varphi^2} + \frac{1}{\r} \frac{\p f}{\p \r}   \quad .
\eea
Note that for the axisymmetric case under consideration in this paper (\ref{cily-diff1})-(\ref{cily-diff3}) simplify further because no function depends on the $\varphi$ angle.\\

The three dimensional cylindrical coordinates $(\r, z, \varphi)$ are related to prolate spheroidal coordinates $(x,y,\z)$ with the following transformation
\bea
         \r(x,y,\z) &=& \l \sqrt{x^2-1} \sqrt{1-y^2} \cos \z  \quad , \\
         z(x,y,\z)  &=& \l x y  \qquad , \\ 
         \varphi(x,y,\z) &=&  \l \sqrt{x^2-1} \sqrt{1-y^2} \sin \z \quad .  \
\eea
The constant $\l$ determines the ellipticity of the coordinates.\\ 
The inverse coordinate transformation is given by
\bea \label{coord-x-y-zetag1}
           x(\r,z,\varphi)  &=&  \frac{1}{2\l} \left[ \sqrt{\r^2 + \varphi^2 + (z+\l)^2} + \sqrt{\r^2 + \varphi^2 + (z - \l)^2} \right]  \quad  , \\
     \label{coord-x-y-zetag2}      y(\r,z,\varphi)  &=&  \frac{1}{2\l} \left[ \sqrt{\r^2 + \varphi^2 + (z+\l)^2} - \sqrt{\r^2 + \varphi^2 + (z - \l)^2} \right]  \quad ,  \\
    \label{coord-x-y-zetag3}       \z(\r,z,\varphi) &=& \arctan \left( \frac{\varphi}{\r} \right) \quad .
\eea
In this set of coordinates the axisymmetry simplification become $\z=0$, giving exactly eq. (\ref{cor-x-y}). In case we want to pass from these to the radial coordinates centred in one of the two sources, (\ref{cor-x-y}) have to be combined with (\ref{barr})-(\ref{theta-cord}), as follows
\beq \left\{\begin{matrix} 
   x &=& \frac{1}{2\l}  \left[ \bar{r} + \sqrt{\bar{r}^2 - 4 \l \bar{r} \cos \theta + 4 \l^2}  \right] \quad , 
   \vspace{0.15cm} \\
   y &=& \frac{1}{2\l}  \left[ \bar{r} - \sqrt{\bar{r}^2 - 4 \l \bar{r} \cos \theta + 4 \l^2}  \right] \quad .        
\end{matrix}\right.
\eeq
Finally the three-dimesional gradient and laplacian in coordinares (\ref{coord-x-y-zetag1})-(\ref{coord-x-y-zetag3}) reads \footnote{In the reference \cite{reina-treves} and in the Carmeli book there are some differences in the form of the differential operators, probably due to typos.}
\bea
   \hspace{-1cm}     \overrightarrow{\nabla} f(x,y,\z) &=& \frac{\overrightarrow{e}_x}{\l} \sqrt{\frac{x^2-1}{x^2-y^2}} \frac{\p f}{\p x} + \frac{\overrightarrow{e}_y}{\l} \sqrt{\frac{1-y^2}{x^2-y^2}} \frac{\p f}{\p y}  + \frac{\overrightarrow{e}_y}{\l \sqrt{x^2-1} \sqrt{1-y^2}} \frac{\p f}{\p \z} \ \ , \\
   \hspace{-1cm}       \nabla^2 f(x,y,\z) &=& \frac{1}{\l^2 (x^2-y^2)} \left\{ \frac{\p}{\p x} \left[ (x^2-1) \frac{\p f}{\p x} \right] + \frac{\p}{\p y} \left[ (1-y^2) \frac{\p f}{\p y} \right] \right\} + \frac{1}{\l^2 (x^2-1)(1-y^2)} \frac{\p^2 f}{\p \z^2} \ \ . 
\eea
\\

\section{Kerr-Newman-NUT solution from the enhanced Ehlers transformation}
\label{app1}

For sake of completeness we sum up the main results of the generalised enhanced Ehlers-Reina-Treves transformation applied to the Kerr-Newman black hole to obtain the Kerr-Newman-NUT solution. \\
In practice we apply, instead of the standard Ehlers transformation (\ref{su21-transf}-III) the enhanced generalised Reina-Treves transforamtion (\ref{gen-enhanced-reina-treves}), to the Kerr-Newman seed ($\Er_0 , \mathbf{\Phi}_0$) of eqs  (\ref{KN-seed-Er0})-(\ref{KN-seed-Phi0}). Thus instead of (\ref{ehlers_Er1})-(\ref{ehlers_Phi1}) we have

\bea \label{ehlers_ErN}
       \Er_0(r,x) \longrightarrow \Er_N(r,x) &=&  \frac{\Er_0 (r,x) + i b}{1+i b\Er_0 (r,x)}  \ = \  1 + \frac{2 m (i+b)}{ax-ir + b(r-2m+iax)}  \quad , \\
       \mathbf{\Phi}_0(r,x) \longrightarrow  \mathbf{\Phi}(r,x) &=&  \frac{\mathbf{\Phi}_0 (r,x) (1+ib)}{1+ib\Er_0 (r,x)} \ = \ \frac{(b-i)(q+ip)}{ax-ir + b(r-2m+iax)}            \quad , \label{ehlers_PhiN}
\eea
thus, thanks to the definitions (\ref{def-Phi-Er}), (\ref{A-tilde}) and (\ref{h}), we obtain 
\bea
      f (r,x) &=& \textsf{Re}(\Er_N) + \mathbf{\Phi}_N \mathbf{\Phi}_N^*   \ = \ \frac{(1+b^2)(p^2+q^2-2mr+r^2+a^2x^2)}{r^2-4abmx+a^2x^2+b^2[(r^2-2m)^2+a^2x^2]} \quad , \quad \\
       A_t (r,x) &=& \textsf{Re}(\mathbf{\Phi}_N) \ = \ \frac{2bmp-qr-apx+b^2(2mq-qr-apx)}{r^2-4abmx+a^2 x^2+b^2[(r^2-2m)^2+a^2x^2]} \quad , \quad  \\
\om(r,x) &=&  -\frac{4bmx}{1+b^2} - \frac{a(1-x^2)\left\{p^2+q^2-2mr+b^2\left[p^2+q^2+2m(r-2m) +4abmx\right]\right\}}{(r^2-2mr+q^2+p^2+a^2x^2)(1+b^2)}  + \om_0 \qquad ,\\
     A_\varphi (r,x) &=&  A_{\varphi 0} + px - \left[ \frac{4mxb}{1+b^2}  + a (1-x^2) - \om_0 \right]  A_t (r,x) \quad .
\eea
Then the standard Kerr-Newman solution (\ref{hatA}), (\ref{kn-nut-standard-inizio}) - (\ref{kn-nut-standard-fine}) can be obtained by the following mass and radial coordinate shift:
\beq \label{bmbar}
 m \longrightarrow \bar{m}=\sqrt{m^2-\ell^2}   \qquad , \qquad \quad  r \longrightarrow \bar{r} = r + \bar{m} -\sqrt{\bar{m}^2+\ell^2}    \quad .
\eeq
In this case $t,a,q,p$ remain the same. The nut parameter $\ell$ is related to the enhanced generalised Reina-Treves transformation parameter $b$ through the equation
\beq
          b =  \frac{-\bar{m}+\sqrt{\bar{m}^2+\ell^2}}{\ell} \qquad .  \label{bmell}
\eeq
As can be easily understood the new transformation is more economical because no extra transformation is required and minimal adjustment of the parameter and coordinates is needed. Moreover it shorten the results, therefore it simplifies also the interpretation of the generated output, in particular when the output is a novel unknown solution.  \\
In order to support the role of the enhanced Ehlers transformation as the gravitational-gravitomagnetic duality, as proposed in section \ref{Ehlers-section},  it is worth also to write explicitly the falloff of the Ernst complex field of the Kerr-Newman solution after the enhanced Ehlers transformation (\ref{gen-enhanced-reina-treves})
\bea
        \Er (r,x) & \sim &  1 - \frac{2 m}{r} \ \frac{i+b}{i-b} - \frac{2 m (i+b) (ax-2bm+iabx)}{(1-b)^2 \ r^2} + O \left( \frac{1}{\bar{r}^3} \right) \ \ , \label{Eexp} \\
       { \bf \Phi } (r,x) & \sim & \frac{-q-ip}{\bar{r}} + \frac{(q+ip)(2bm-ax-iabx)}{(i-b) \ r^2}  + O \left( \frac{1}{\bar{r}^3} \right) \ \  . \label{PHIexp}
\eea
We can understand from eq. (\ref{Eexp}) that, if the seed spacetime is not endowed with mass charge, the (enanched) Ehlers transformation is not adding any nut charge, as noted in section \ref{Ehlers-section}. This occurs because the role of (enanched) Ehlers transformation is just to rotate the gravitational mass charge into the gravomagnetic charge, in the same way the electromagnetic duality rotates the electric monopole charge into the magnetic charge (or viceversa), but the latter is not generating any magnetic charge from electric uncharged seed. A further advantage of the enhanced Ehlers transformation, with respect to the traditional one, is revealed by this duality. Indeed for the standard Ehlers transformation the gravitational duality is not clear because it mixes also the electric with the magnetic field and because non-trivial coordinate transformations are needed to obtain the desidered spacetime.   \\
Thanks to the parameter rescaling\footnote{The radial shift in (\ref{bmbar}) is not relevant, it is just changing the geometrical constant $z*$ of the Ernst potential expansion (\ref{complex-asym}) - (\ref{complex-asym2}), which is related to the set of coordinate under consideration.} (\ref{bmbar}) we get the standard (dyonic) Kerr-Newman-NUT asymptotic 
for the complex fields
\bea
        \Er(\bar{r},x) & \sim &  1 - 2 \ \frac{\bar{m}-i\ell}{\bar{r}}  - \frac{2(\bar{m} - i\ell)\left( -iax+i\ell -2\bar{m} +\sqrt{m^2+\ell^2}\right)}{\bar{r}^2} + O \left( \frac{1}{\bar{r}^3} \right) \ \ ,  \\
       { \bf \Phi } (\bar{r},x) & \sim & \frac{-q-ip}{\bar{r}} + \frac{(p-iq)\left( - a x + \ell +2i\bar{m} -i \sqrt{\bar{m}^2+\ell^2} \right)}{\bar{r}^2} + O \left( \frac{1}{\bar{r}^3} \right) \ \ ,
\eea
From the latter expansion it is straightforward to identify the conserved quantities of the solution: the mass $\bar{m}$, nut charge $\ell$, angular momentum $J=am$, electric and magnetic charges $-q$ and $-p$ respectively, electric and magnetic dipole $D_e=-ap$ and $D_q=aq$.
\\
Note that the complete rotation of the mass charge into the gravitomagnetic one occurs for b=1. In the example considered in this appendix it means $\bar{m}=0$ or $m=\ell$. \\

\subsection{Multipole moments expansion for the Kerr-NUT spacetime}
\label{app-multipole}

It is clarifying to analyse the mass and the angular multipolar expansion for a metric possessing both angular momentum and the NUT parameter in order to understand the contribution of the NUT parameter to the rotation of the spacetime. Probably the easiest example is just given by the above Kerr-NUT metric, so we can consider the electromagnetic charges $q$ and $p$ null in this subsection.\\
It is possible to compute the multipoles expansion directly from the Ernst potential (\ref{ehlers_ErN}), as explained in \cite{quevedo}. The first four orders of the mass and angular multipoles are respectively given by the following formulas

\bea \label{polesM}
           M_l &=& \textsf{Re} \left[  \frac{1}{l!}\frac{d \tilde{\xi}(\tilde{z},1)}{d\tilde{z}} \bigg|_{\tilde{z}=0} \right] \quad , \\
     \label{polesJ}      J_l  &=& \textsf{Im}\left[  \frac{1}{l!}\frac{d \tilde{\xi}(\tilde{z},1)}{d\tilde{z}}  \bigg|_{\tilde{z}=0} \right]\quad ,
\eea

where $\tilde{\xi}(\tilde{z},1)=\frac{1}{\tilde{z}} \xi(\tilde{z},1)$ can be inferred, after a coordinate transformation $r \rightarrow \frac{1}{\tilde{z}} + M$ which locates the spacelike infinity at $\tilde{z}=0$, from the gravitational Ernst potential (\ref{ehlers_ErN}) evaluated on the azimuthal symmetry axis $x=1$

\beq 
        \xi(\tilde{z},1)= \frac{1-\Er_N(\tilde{z},1)}{1+\Er_N(\tilde{z},1)} \quad .
\eeq
\\
Let us write explicitly, once we have defined $\s := \sqrt{M^2+\ell^2} $, the first four orders for the Kerr-NUT black hole:

\bea
       M_0 &=& M \quad , \nn \\ 
       M_1 &=& -M^2 -a\ell + M \s \quad , \nn \\
       M_2 &=& - a^2 M + 2 a \ell \left( M - \s  \right) + M \left[ \ell^2 + 2 M \left( M - \s \right) \right] \quad ,   \qquad         \nn  \\ 
       M_3 &=& a^3\ell + 3a^2M\left(M - \s  \right)+ M\left(-4 M^3-3M\ell^2+4M^2 \s  + \ell^2\s \right) -3a\ell \left[\ell^2 + 2 M \left( M - \s  \right) \s  \right] \quad , \qquad     \nn  \\
       J_0 &=& \ell \quad , \label{kerr-nut-poles} \\
       J_1 &=& - a M + \ell \left( M - \s  \right) \ \ ,  \nn \\
       J_2 &=& (a-\ell) \left[ \ell (a+\ell) +2M\left(M - \s  \right) \right] \quad , \nn \\
       J_3 &=& a^3 M + 4   M^3 \ell + 3 M \ell^3 -4M^2\ell\s - \ell^3\s +3a^2\ell\left(\s-M \right) -3aM\left[ \ell^2+2M\left( M - \s \right) \right]   \quad   .     \nn  
\eea

When the NUT parameter vanish, i.e. $ \ell \rightarrow 0$, these values simplify to the Kerr multipole moment expansion

\bea
             M_0 &=& M \quad , \qquad M_1 = 0 \ \ \qquad , \qquad M_2 = - a^2 M \quad , \qquad M_3 = 0 \qquad \ \ , \qquad   \\
             J_0 &=& 0 \ \ \quad , \  \qquad J_1 = - a M \quad , \qquad \ J_2 = 0 \quad \  \qquad , \qquad \ J_3 = a^3 M \quad .
\eea 

From the above example we can see as the NUT parameter $\ell$  contribute to the multipole moments, in particular it represent the angular monopole moment of the source. It is clear that also in the case of null Kerr rotational parameter $a$, the resulting Taub-NUT spacetime is not static but is stationary rotating, even though its angular momentum (proportional to the angular dipole moment $J_1$) is null, as can be seen from the $ a \rightarrow 0 $ limit of the expansion (\ref{kerr-nut-poles})

\bea
       M_0 &=& M \quad , \nn \\ 
       M_1 &=& -M^2 + M \s \quad , \nn \\
       M_2 &=& M \left[ \ell^2 + 2 M \left( M - \s \right) \right] \quad ,   \qquad         \nn  \\ 
       M_3 &=& M\left(-4 M^3-3M\ell^2+4M^2 \s  + \ell^2\s \right) \quad , \qquad     \nn  \\
       J_0 &=& \ell \quad , \label{taub-nut-poles} \\
       J_1 &=& \ell \left( M - \s  \right) \ \ ,  \nn \\
       J_2 &=& -\ell \left[ \ell^2 +2M\left(M - \s  \right) \right] \quad , \nn \\
       J_3 &=&  4 M^3 \ell + 3 M \ell^3 -4M^2\ell\s - \ell^3\s    \quad   .     \nn  
\eea

Moreover the NUT parameter $\ell$ switches on all the even multipole moments which otherwise would have been null.\\ 
For the double Reissner-Nordstrom-NUT spacetime, both in the extremal and non-extremal case, build in section \ref{sec-majumbdar} the NUT parameter has a similar behaviour as can be read directly from the asymptotic expansion of the complex Ernst potentials (\ref{E-2RN})-(\ref{Phi-2RN}), which capture the first orders of the multipole moments, as already commented in section \ref{alekseev-beli-nut}. 
Therefore the double Reissner-Nordstrom-NUT spacetime posses a kind of stationary rotation even though its angular momentum is null, which, similarly to the standard angular momentum, is able to twist geodesics.\\

\section{NUT generalisation of the Majumdar-Papapetrou in $(\r,z)$ coordinates}
\label{app-rho-z}

It is not difficult to get the NUT generalisation of the Majumdar-Papapetrou metric in the original cylindrical coordinates ($\rho,z$) of the Lewis-Wayl-Papapetrou metric (\ref{lwp-metric})
\beq 
                         ds^2 = - f \left( dt + \om d\varphi \right)^2 + f^{-1} \left[ \rho^2 d\varphi^2 + e^{2\gamma}  \left( d \rho^2 + d z^2 \right) \right] \ .
\eeq
One has just to transform the functions given in (\ref{fxy})-(\ref{magnbxy}) according with the coordinates transformation (\ref{cor-x-y}) to get
\bea
        f^{-1}(\r,\theta) &=&  \left[ \frac{1-b^2}{1+b^2} +  \frac{m_1}{\sqrt{\r^2 + (z+\l)^2}} + \frac{m_2}{\sqrt{\r^2 + (z-\l)^2}} \right]^2 + \left( \frac{2 b}{1+b^2}  \right)^2    \quad , \\
        \omega(\r,\theta) &=& \frac{4 b}{1+b^2}  \left[ \frac{m_1 (z+\l)}{\sqrt{\r^2 + (z+\l)^2}} + \frac{m_2 (z-\l)}{\sqrt{\r^2 + (z-\l)^2}} \right] + \omega_0     \quad , \nn \\
        \gamma(\r,\theta) &=& 0      \quad , \nn \\
         A_t(\r,\theta) &=&  f(\r,\theta) \left\lbrace  \left( \frac{1-b^2}{1+b^2}\right) \left[ \frac{1}{2}  \left( \frac{1 + b^2}{1 - b^2}\right) + \frac{m_1}{\sqrt{\r^2 + (z+\l)^2}} + \frac{m_2}{\sqrt{\r^2 + (z-\l)^2}} \right]^2 -\frac{1}{4}  \left( \frac{1 + b^2}{1 - b^2}\right)  \right\rbrace                          \quad , \nn \\
         A_\varphi(\r,\theta) &=&   \omega(\r,\theta) A_t(\r,\theta) + A_{\varphi_0}                           \quad .
\eea
In this form the solution is not only more compact, but also it is more suitable to describe the near horizon geometry of the two sources simultaneously, as done in (\ref{nh-m1m2}). \\
\\

\end{document}